\documentclass{WileyMSP-template}

\usepackage{graphicx}
\usepackage{color}
\usepackage{colordvi}  
\usepackage{amsmath}
\usepackage{amssymb}
\usepackage{bm}
\usepackage{cancel}
\usepackage{soul}
\usepackage{amsfonts}

\usepackage{multicol}
\usepackage{lipsum}

\usepackage[colorlinks=true,linkcolor=blue]{hyperref}

\newcommand{\tbeta}{\tilde{\beta}}

\newcommand{\bLambda}{{\bm \Lambda}}

\newcommand{\bphi}{{\bm \phi}}

\newcommand{\balpha}{{\bm \alpha}}

\begin{document}

\pagestyle{fancy}

\title{Bloch--Landau--Zener oscillations in moir\'e lattices}

\maketitle


\author{Sergey~K.~Ivanov$^1$*, Yaroslav~V.~Kartashov$^2$, and Vladimir~V.~Konotop$^3$}

\dedication{ 
	$^1$Instituto de Ciencia de los Materiales, Universidad de Valencia, Catedr\'{a}tico J. Beltr\'{a}n, 2, Paterna, 46980, Spain
	}

\dedication{ 
	$^2$Institute of Spectroscopy, Russian Academy of Sciences, Troitsk, Moscow 108840, Russia 
}

\dedication{ 
$^3$Departamento de F\'isica and Centro de Física Te\'orica e Computacional, Faculdadede de Ci\^encias da Universidade de Lisboa, Campo Grande, Ed. C8, Lisboa 1749-016, Portugal}

\dedication{ 
	$^*$sergei.ivanov@uv.es}




\begin{affiliations}

~


~


\end{affiliations}



\begin{abstract}

\justifying
We develop a theory of two-dimensional Bloch–Landau–Zener (BLZ) oscillations of wavepackets in incommensurate moir\'e lattices under the influence of a weak linear gradient. Unlike periodic systems, aperiodic lattices lack translational symmetry and therefore do not exhibit a conventional band-gap structure. Instead, they feature a mobility edge, above which (in the optical context) all modes become localized. When a linear gradient is applied to a moir\'e lattice, it enables energy transfer between two or several localized modes, leading to the oscillatory behavior referred to as BLZ oscillations. This phenomenon represents simultaneous tunneling in real space and propagation constant (energy) space, and it arises when quasi-resonance condition for propagation constants and spatial proximity of interacting modes (together constituting a selection rule) are met. The selection rule is controlled by the linear gradient, whose amplitude and direction play a crucial role in determining the coupling pathways and the resulting dynamics. We derive a multimode model describing BLZ oscillations in the linear regime and analyze how both attractive and repulsive nonlinearities affect their dynamics. The proposed framework can be readily extended to other physical systems, including cold atoms and Bose–Einstein condensates in aperiodic potentials.

\end{abstract}


\twocolumn
\justifying

\section{Introduction}

When a quantum particle in a periodic potential is subjected to a weak linear force, Bloch oscillations~\cite{Bloch1929} can be observed. This fundamental phenomenon can be interpreted as the adiabatic motion of the particle in the reciprocal space, during which it follows the periodic dispersion relation representative for the band-gap spectrum of periodic potentials. Being ubiquitous across different areas of physics, Bloch oscillations have been experimentally observed in condensed matter~\cite{Mendez1988, FeLeSh, Leo1992, Waschke1993}, atomic~\cite{Dahan1996, Wilkinson1996, AndKas, Morsch2001, Cristiani2002, Ferrari2006, Gustavsson2008, Kling2010, Geiger2018}, and diverse optical~\cite{Peschel1998, Pertsch1999, Morandotti1999, Sapienza2003, Dreisow2009, Yuan2016} systems.  

In recent years, there has been a growing interest in the investigation of various dynamical phenomena in structured \textit{aperiodic} potentials, conducted using setups similar to those employed for periodic media. For the one-dimensional (1D) quasi-periodic potentials, prominent examples include the evolution of atoms in optical lattices \cite{Reeves2014, Luschen2018}, nontrivial beam dynamics in arrays of optical waveguides \cite{Kraus2012} and induced lattices in photosensitive materials \cite{Yang2024}. A natural question that arises is: what would be the dynamics of a wavepacket in such quasi-periodic systems under the influence of a weak gradient? When a quasi-periodic potential is formed by two sublattices of substantially different depths, such that one of the sublattices can be treated as a perturbation, the tight-binding approximation can be applied to the deeper lattice, giving rise to discrete models where the incommensurate lattice acts as a modulation. This approach leads to the well-known Aubry-Andr\'e model~\cite{Aubri1980}. The study of Bloch oscillations induced by an additional weak external field in such 1D aperiodic systems has been reported in Refs.~\cite{Moura2005, Wang2014}. In the fully continuous 1D models with incommensurate lattices, damping of Bloch oscillations was explored in~\cite{Walter2010},  where the aperiodic lattice was also approximated as a deep periodic potential with a weak incommensurate perturbation. Experimental studies of atomic Bose-Einstein condensates (BECs) in a bichromatic lattice with incommensurate periods were performed in~\cite{Reeves2014}, where a reduction of damping due to interatomic interactions was observed. The theory of Bloch–Landau–Zener (BLZ) oscillations of localized modes in 1D quasiperiodic lattices with comparable potential depths was developed in~\cite{Prates2024}.

If a quasi-periodic potential is generated by two (or more) incommensurate lattices of comparable amplitudes, neither the tight-binding approach nor the perturbative treatment of one of the constituent sublattices is applicable. In this case, the Hamiltonian spectrum no longer exhibits a band-gap structure, but instead acquires fractal characteristics~\cite{Simon1982}, thereby losing one of the key features of periodic systems typically employed to interpret Bloch oscillations. Furthermore, when the potential is not shallow, two main types of eigenstates emerge: spatially localized and extended ones, separated by a mobility edge (ME)~\cite{Sarnak1982, Kohmoto83, Surace1990, Diener01, Modugno2009, Biddle2002} (even more general situations where several MEs, as well as critical states can exist, will not be addressed here). Nevertheless, even under these conditions, a weak linear gradient can still induce oscillatory dynamics, although the nature of these oscillations differs from those in periodic systems. For the 1D localized states, it was shown in Ref.~\cite{Prates2024} that the oscillatory dynamics arises from tunneling among \mbox{(quasi-)}\hspace{0pt}resonant eigenstates, and in this sense it closely resembles the phenomenon of Landau-Zener tunneling~\cite{Zener1932, LandauTun1932}, {which has been experimentally studied in systems of cold atoms loaded in periodic lattices~\cite{Sias2007,Wilkinson1997,Zenesini2008,Zenesini2009,Tayebirad2010}.} Consequently, the resulting dynamics have been termed Bloch-Landau-Zener (BLZ) oscillations, the term that will be used below.

The physics underlying Bloch oscillations in 1D periodic structures and BLZ dynamics in 1D quasi-periodic systems differ qualitatively, resulting in mar\-ked\-ly distinct dynamical behaviors~\cite{Prates2024}.  For instance, the amplitude and frequency of BLZ oscillations depend on both the initial conditions and the applied gradient. Unlike standard Bloch oscillations, BLZ behavior can involve more than two eigenstates and may feature multiple characteristic frequencies of oscillations. Increasing linear gradient may even suppress BLZ oscillations --- a phenomenon that never occurs in periodic systems. Notably, BLZ oscillations also exhibit a remarkable robustness against weak nonlinearities of both types, attractive and repulsive, in contrast to conventional Bloch oscillations.

Bloch oscillations in 2D periodic lattices {(continuous, discrete, and semi-discrete)} have also been considered in earlier theoretical studies~\cite{Gluck2001, Kolovsky2003, Witthaut2004, Mossmann2005, Khomeriki2010, Kolovsky2011,Kolovsky2012_1,Kolovsky2012_2,Kolovsky2013,Maksimov2015,Kolovsky2016,Driben2017, Ye2023}, and observed experimentally in optically induced lattices~\cite{Trompeter2006} and in phononic crystals~\cite{He2007}. The dynamics of such oscillations were touched in~\cite{Moura2008}, where a sophisticated aperiodic potential was modeled using tight-binding approximation, as well as in the bilayer discrete moir\'e lattice~\cite{Yar2023}, where localized modes were not considered. Generally, BLZ dynamics in continuous 2D aperiodic systems remain largely unexplored, the physical mechanism has not been described, and the modes participating in these processes have not been identified.

The goal of this paper is to extend the theory of BLZ oscillations to 2D aperiodic continuous systems, focusing on the propagation of light beams through photonic moir\'e arrays. Although the analysis is carried out in the context of photonic systems, the underlying model is broadly applicable, e.g., to cold atoms and BECs in moir\'e potentials. There are several features of the 2D aperiodic system with gradient, making it qualitatively different from 1D aperiodic systems. These include eventual level crossing of localized states, a much larger variety of the eigenstates of the governing Hamiltonian (say, including states localized along only one spatial direction), coupling scenarios, as well as the presence of additional control parameters like the twist angle of a moir\'e lattice and the orientation of the applied weak gradient. Nonlinearity in 2D systems can lead to a wider range of instabilities, including collapse or enhanced diffraction.  

 BLZ oscillations described in our work are readily observable experimentally, as photonic moiré lattices can be created using well-established methods. These include photorefractive crystals~\cite{Wang2020, Fu2020, Wang2022}, femtosecond-laser written waveguide arrays~\cite{Arkhipova2023}, and photonic crystals with moiré patterns produced via holographic fabrication techniques~\cite{Hurley2023}. The holographic method also enables the creation of a gradient in photonic moir\'e lattices through the addition of an auxiliary sublattice~\cite{Shang2021}. A refractive index gradient can also be induced in waveguide arrays via thermo-optic response by applying a thermal gradient to a sample \cite{Toyoda1983,Plotnik2011}.

The paper is organized as follows. Section~2 introduces the problem and describes the system under consideration. In Section~3, we derive the selection rule governing BLZ oscillations, which involves conditions of spatial proximity and quasi-resonance between the participating states. This section also presents examples of periodic energy transfer between such states. Section~4 examines the impact of focusing and defocusing nonlinearities on BLZ dynamics. {Finally, in Section~5, we discuss several aspects related to the experimental feasibility of observing 2D BLZ oscillations.}

\section{Statement of the problem}

We consider the propagation of a paraxial optical beam along the $z$ direction in the material with shallow transverse modulation of the refractive index, governed by the dimensionless Schr\"odinger equation for the light field amplitude $\Psi$ (for normalizations see \cite{Fu2020}):
\begin{eqnarray}
	\label{main}
	i\frac{\partial \Psi}{\partial z}=H_{\balpha}\Psi ,
\end{eqnarray}
where the Hamiltonian is given by
\begin{eqnarray}
\label{Hamilt}
	H_{\balpha}= -\frac{1}{2}\nabla^2-U({\bm r})-\balpha\cdot{\bm r} ,  
\end{eqnarray}
where ${\bm r}=(x,y)$ and $\nabla=(\partial_x,\partial_y)$. The optical potential $U({\bm r})$ represents an incommensurate (i.e., aperiodic) moir\'e lattice \cite{Wang2020, Fu2020, Wang2022} induced by two sublattices. These sublattices are mutually rotated by a non-Pythagorean angle $\theta=\pi/6$, implemented via the 2D rotation operator $R(\theta)$, and shifted relative each other by the displacement vector $\bm s$:
\begin{eqnarray}
	\label{pot_total}
	 U({\bm r})=p_1V(R(\theta){\bm r}+\bm{s})+p_2V({\bm r}).
\end{eqnarray}
Here $p_{1,2}$ are the depths of the sublattices (it is assumed that $|V({\bm r})|\leq 1$). The linear gradient is characterized by the vector $\balpha=(\alpha_x,\alpha_y)$, which is considered small enough, i.e., $\alpha=|\balpha|\ll p_{1,2}$. The shift $\bm{s}$ is introduced to break the discrete rotational symmetry of the lattice. Neither the choice of a non-Pythagorean angle, nor the shift $\bm s$ are essential for the consideration, provided the respective lattice is incommensurate phase (say, $\bm{s}$ can acquire zero value).  

The respective stationary eigenvalue problem, obtained by using the ansatz $\Psi=e^{ibz}\varphi({\bm r})$, where $b$ is the propagation constant, has the form 
\begin{eqnarray}
	\label{Halpha}
	H_{\balpha}\varphi_{\bm{n}}^{(\balpha)}=-b_{\bm{n}}^{(\balpha)} \varphi_{\bm{n}}^{(\balpha)}.
\end{eqnarray}
We adopt a vector notation for labeling the eigenstates $\bm{n}=(n_1,n_2)$ to account for possible degeneracies of the eigenvalues $b_{\bm{n}}^{(\balpha)}$. Such degeneracies may originate from discrete rotational symmetry (if $\bm{s}=(0,0)$) or occur accidentally [see, e.g., Fig.~\ref{fig2} below]. The first index $n_1$ numbers the eigenvalues in the descending order of their absolute values, while the second index $n_2$ numbers eigenfunctions within the respective degenerate subspace. We also explicitly indicate the dependence of the eigenfunctions on the gradient $\bm{\alpha}$. In the absence of the gradient, i.e., for $\balpha=(0,0)$, one obtains the eigenvalue problem
\begin{eqnarray}
H_0\varphi_{\bm{n}}^{(0)}=-b_{\bm{n}}^{(0)} \varphi_{\bm{n}}^{(0)}, 
\quad H_0=-\frac{1}{2}\nabla^2-U({\bm r})	 ,
\end{eqnarray} 
which describes modes of a conventional aperiodic moir\'e lattice.

Strictly speaking, the spectrum of the Hamiltonian $H_{\balpha}$ with $\alpha>0$, considered in the entire space $\mathbb{R}^2$, is expected to be continuous, implying the absence of square-integrable eigenstates (see, e.g.,~\cite{Avron1977,Bentosela1983,Nenciu1991} for rigorous results in 1D). Nevertheless, it is relevant from both experimental and theoretical perspectives to consider a moir\'e lattice with a large, but finite area $S$, with zero boundary conditions for eigenmodes. In such a statement, one deals with a discrete spectrum of $H_{\balpha}$, which justifies the use of a discrete two-component index $\bm n$. On the other hand, as we will see below from a thorough numerical analysis, the moir\'e lattice with nonzero $\balpha$ supports spatially localized states, for which the choice of the specific boundary conditions is not relevant for observed dynamics. The respective eigenmodes can be normalized:
\begin{eqnarray}	\int_{S}|\varphi_{\bm{n}}^{(\balpha)}|^2d{\bm r}=1,
\end{eqnarray}
where the integration is carried out over the whole lattice area $S$.

As it was shown in previous numerical and experimental studies of moir\'e lattices (see e.g.~\cite{Wang2020,Fu2020,Wang2022,Huang2016}), when the lattice is sufficiently deep, the eigenmodes of $H_0$ with largest eigenvalues $b_{\bm{n}}^{(0)}$ (the lower states in quantum-mechanical applications) become spatially localized. The indicator of localization of such modes is their form-factor $[\chi_{\bm{n}}^{(\balpha)}]^{1/2}$ where
\begin{eqnarray}
\chi_{\bm{n}}^{(\balpha)}=\langle[\varphi_{\bm n}^{(\balpha)}]^2, [\varphi_{\bm n}^{(\balpha)}]^2\rangle 
\end{eqnarray}
(it is also known as the inverse participation ratio), and  $\langle f,g\rangle :=\int_Sf^*gd{\bm r}$. The larger the value of $\chi$, the stronger the localization. In particular, modes with the largest eigenvalues tend to localize within a small region of the total domain, such that
\begin{eqnarray}
	\chi_{\bm{n}}^{-1}  \ll S.
\end{eqnarray}

The parameters of the lattice, at which localized modes emerge for a given gradient, correspond to the localization-delocalization transition (LDT). Increase of the lattice depth above this transition leads (in a general situation) to an increase in the number of localized modes. Typically, when the parameters are above the LDT, one can identify a mobility edge, $b_{\rm ME}^{(\bm\alpha)}$, which can be defined as the largest propagation constant of an extended state, such that all modes with $b_{\bm n}^{(\bm\alpha)}>b_{\rm ME}^{(\bm\alpha)}$ in the area $S$ are localized. Although this is not a strict definition, the latter requires information about the modes defined in the entire space $\mathbb{R}^2$, it is practically useful for characterizing BLZ oscillations, the focus of this work.

\begin{figure}[t]
\centering
\includegraphics[width=1.00\linewidth]{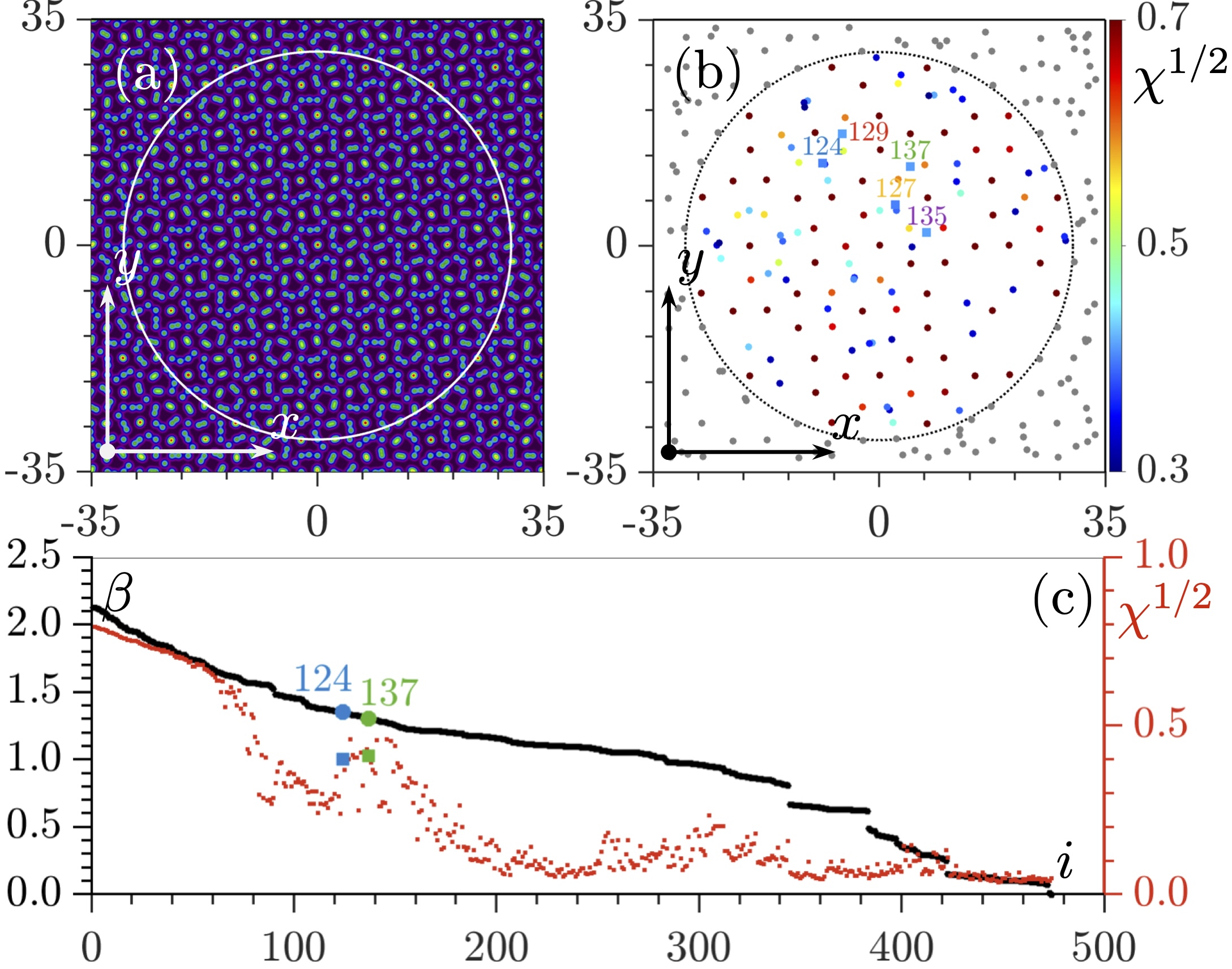}
\caption{(a) Profile of the moir\'e lattice for $\bm\alpha=0$. (b) Positions of the centers of mass ${\bm r}_{i}^{(0)}={\bm r}_{\bm{n}}^{(0)}$ of different localized modes with $\chi\ge 0.3$, corresponding to the ordering $\bm n\to i$ described in the text, for the lattice without gradient. The color code is used to indicate the form-factors $\chi^{1/2}$ of corresponding modes. The indices $i$ of the modes used below for dynamical simulations are highlighted by the numbers. The white circle in panel (a) and the dotted black circle in panel (b) with radius $\rho=30$ indicate the cutoff used to restrict the number of modes in the subsequent numerical analysis of the dynamics. This ensures that only modes sufficiently far from the boundaries of the simulation domain are included. (c) Propagation constants $\beta$ plotted as a function of mode index $i$ for $\bm\alpha=0$ (black circles). The corresponding squared form-factors $\chi$ are shown as red squares (right vertical axis). Blue and green labels indicate the specific modes chosen as initial conditions for dynamical simulations. Here and in all figures below, $p_1=p_2=4$, $d=2.5$, $w=0.5$, and $\theta=\pi/6$.
}
\label{fig1}
\end{figure}

The lattice under consideration and the properties of its highest eigenmodes are illustrated in Fig.~\ref{fig1}. Fig.~\ref{fig1}(a) shows the lattice profile within a large but finite domain $S$, which is used in all simulations presented below. We consider an experimentally relevant configuration in which each sublattice $V$ is composed of Gaussian waveguides arranged in a square array:
\begin{eqnarray}
	\label{Gaussain}
	V({\bm r})=-\sum_{{\bm m}\in\mathbb{Q}^2} e^{-({\bm r}-\bm{m}d)^2/w^2},
\end{eqnarray}
where $\mathbb{Q}^2$ is the space of 2D vectors with integer components defining the Bravais lattice with period $d$ and $w$ is the waveguide radius. Two sublattices are rotated by the angle $\theta=\pi/6$ corresponding to the incommensurate (aperiodic) moir\'e structure.

Figure~\ref{fig1}(b) shows positions of localized modes emerging in the lattice without tilt, $\bm\alpha=0$, characterized by ``centers of mass'' coordinates defined by
\begin{eqnarray}
\label{COM_mode}
   {\bm r}_{\bm{n}}^{(\balpha)}= \langle \varphi_{\bm{n}}^{(\balpha)}, {\bm r}\varphi_{\bm{n}}^{(\balpha)}\rangle.
\end{eqnarray}
The degree of localization (form-factor) of different modes in this figure is indicated by the color of the corresponding dot. We observe a nearly homogeneous, although irregular, distribution of the localized modes over the whole area of the moir\'e lattice (this is similar to the behavior of modes in 1D quasi-periodic arrays \cite{Prates2024,Prates2022}). Considering this fact, in panels (a) and (b) of  Fig.~\ref{fig1} we outline a central area delimited by the circumference of the radius $\rho=30$, shown as a white solid line in (a) and a black dotted line in (b), inside which the modes used for dynamical simulations are located (this constraint ensures that the boundaries do not affect the dynamics, but it does not have any physical significance, as this area can be further expanded by increasing the integration window). Below, we refer to this domain as the {\em effective area}, denoting it as $S_{\rm eff}$. It is worth emphasizing that its area is  $S= 4900$, while the smallest form-factor shown in Fig.~\ref{fig1}(b) corresponds to  $1/\chi\approx 11.11$, i.e., these modes are well localized within this area (specific examples are shown in the figures below).

Close inspection of the localization properties of the modes reveals that in incommensurate moir\'e lattices, in contrast to 1D quasi-periodic lattices, the ME is not sharp. Instead, delocalization of modes occurs relatively smoothly as their propagation constant varies. This is illustrated in Fig.~\ref{fig1}(c), where black dots represent the propagation constants $\beta$ of the modes confined within the effective area $S_{\rm eff}$, and red squares show corresponding form-factors $\chi^{1/2}$ for each mode (the notation $\beta$ for the propagation constant will be explained below). There are no large gaps in the propagation constant spectrum, to which the ME could belong, which also contrasts with the case of 1D quasi-periodic potentials. Roughly speaking, the 200 modes with the largest propagation constants within the effective area can be considered as well localized. The consideration below will be restricted, namely, to these modes. It is important to stress that we consider here only the eigenmodes of $H_0$ that remain localized in the lattice with gradient described by $H_{\bm \alpha}$ (recall that $\alpha\ll 1$).

Thus, the main goal of this paper is to address the following fundamental question: {\em How does an eigenmode of the Hamiltonian  $H_0$ without a gradient ($\alpha=0$), considered as an initial condition for Eq.~(\ref{main}), evolve under influence of a weak gradient ($0<\alpha\ll 1$)?} For a linear system, this directly describes the evolution of an initial input beam under the influence of the gradient term ${\bm r}\cdot\balpha$.

\begin{figure*}[t]
\centering
\includegraphics[width=1.00\linewidth]{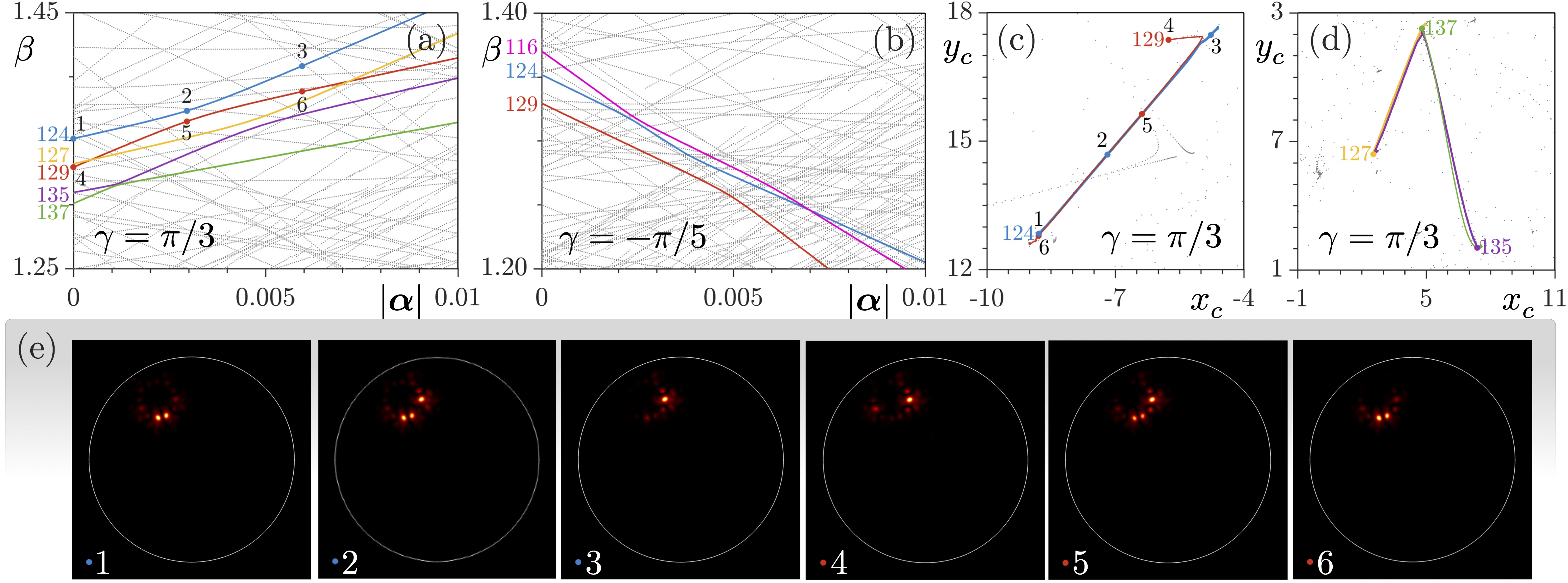}
\caption{(a, b) Propagation constants as functions of $|\balpha|$ for the angles $\gamma=\pi/3$ (a) and $\gamma=-\pi/5$ (b). The branches used below to illustrate the evolution are highlighted in color. (c, d) Positions of the centers of mass ${\bm r}_i^{(\balpha)}=\left(x_c (\alpha),y_c (\alpha)\right)$ of modes $124$ and $129$ (c) and modes $127$, $135$, and $137$ (d) as a function of $\alpha$ for $\gamma=\pi/3$. (e) Mode profiles {$|\phi_{124}^{(\balpha)}|$ and $|\phi_{129}^{(\balpha)}|$ } corresponding to the points marked by numbers 1 to 6 in panels (a) and (c). }
\label{fig2}
\end{figure*}

\section{Multi-mode dynamics}

\subsection{Localized bases of eigenfunctions}

In the 2D case, the Hamiltonians $H_0$ and $H_{\bm \alpha}$ may exhibit degeneracies in their spectra. Moreover, even when there are no degeneracies in the spectrum of $H_0$ and all modes can be ordered by the magnitude of their eigenvalues at $\bm \alpha=0$, there is no guarantee that this ordering will be preserved at nonzero $\bm \alpha$ [see the examples in Fig.~\ref{fig2}(a) and (b)]. To avoid this ambiguity, we introduce a more convenient labeling of localized modes via a single scalar index $i$. Specifically, we define $\beta_i^{(0)}=b_{\bm{n}}^{(0)}$ and set $i_1 < i_2$ if $b_{\bm{n}_1}^{(0)}>b_{\bm{n}_2}^{(0)}$ (respectively $\beta_{i_1}^{(0)}>\beta_{i_2}^{(0)}$). Note that according to the physical meaning of the vector index $\bm n$ described above, if no degeneracies occur, then one has a simple relation $i=n_1$. Within degenerate subspaces (i.e., for modes with identical eigenvalues), the order of indices can be chosen arbitrarily. The resulting set of propagation constants $\beta_i^{(0)}$ is shown in Fig.~\ref{fig1}(c) for $\alpha=0$ as black dots. The corresponding eigenfunctions and their centers of mass are denoted as $\varphi_{\bm{n}}^{(0)}=\phi_{i}^{(0)}$ and ${\bm r}_{\bm{n}}^{(0)}={\bm r}_{i}^{(0)}$, respectively. The index $i$ defined in this way is used in Figs.~\ref{fig1}(b) and (c) to label several representative modes employed below to illustrate different dynamical regimes. Importantly, we restrict this labeling to the modes located within the effective area delimited by the circumference, thereby identifying only the modes that are relevant (employed) in subsequent dynamics. This restriction, however, does not imply any theoretical limitation or ambiguity. Indeed, if the integration domain $S$ is enlarged to $S'$ or even goes to infinity, many additional modes with eigenvalues above the ME will appear. These new modes may alter the global indexing, but they are typically localized outside the original domain $S$, i.e., in $S' \setminus S$. As a result, they do not affect the dynamics of the modes confined well within $S$. Even potential deformations of modes near the boundary $\partial S$ have negligible influence on the interior states. This behavior is consistent with previous observations in 2D systems (e.g., \cite{Wang2020}) and is discussed in greater detail for the 1D case in \cite{Prates2024, Prates2022, Konotop2024}. Thus, increasing the integration domain (assuming it is initially large enough) only affects the absolute numbering of the modes, without altering their physical properties, spatial positions, or ordering of eigenvalues within the region of interest.

We emphasize that the ordering of $\beta_i^{(\balpha)}$ at a given $\balpha$ can differ from the one introduced for  $\beta_i^{(0)}$ due to possible level crossing, i.e., it may happen that $\beta_{{i}_1}^{(\balpha)}<\beta_{{i}_2}^{(\balpha)}$ (or $b_{\bm{n}_1}^{(\balpha)}<b_{\bm{n}_2}^{(\balpha)}$) for $i_1<i_2$.  An example is provided by the yellow and red branches in Fig.~\ref{fig2}(a), which display the dependence of the propagation constants on $\alpha$. These two branches undergo two level crossings within the shown interval of $\alpha$. The yellow branch originates from the state with index 127, while the red branch originates from mode 129. Accordingly, in our labeling scheme, we associate all propagation constants and modes along the yellow branch with index 127, and those along the red branch with index 129, for all values of $\alpha$. Numerous additional level crossings can also be observed in the gray branches.

Let $N$ be the number of the localized modes of $H_0$ (already taking the degeneracy into account) which remain localized in the presence of a nonzero gradient. Considering increasing gradient $(0,0)\to\balpha$ as a deformation, one can trace the result of this deformation upon which localized mode $\phi_{i}^{(0)}$, where $i=1,...,N$, is transformed into the eigenmode $\phi_{i}^{(\balpha)}$ of $H_{\balpha}$. Figure~\ref{fig2}(a) shows the dependence of $b_i^{(\balpha)}$ on $\alpha$ for several modes with propagation constants in the range $1.25$--$1.45$. Further, we assume that the gradient in polar coordinates is characterized by the angle $\gamma$ and strength $\alpha$, i.e., $\balpha = (\alpha \cos\gamma, \alpha \sin\gamma)$, and Fig.~\ref{fig2}(a) corresponds to the case of $\gamma = \pi/3$. Several representative curves in this figure, highlighted in color, correspond to the modes for which the dynamics will be analyzed below. Notably, the modes with indices $124$ (blue branch) and $129$ (red branch) exhibit an avoided crossing near $\alpha \approx 0.003$. In this region, resonant interaction is expected if either of these modes $\phi_{124}^{(0)}$ or $\phi_{129}^{(0)}$ is used as the initial condition for propagation in the system with $\alpha \approx 0.003$. As we will show below, the resonant interaction of two (or more) modes requires yet another condition, namely, a non-negligible coupling of these modes.  For the selected modes, this condition is generally not met even at their level crossing with other modes within the effective area. Figure \ref{fig2}(e) illustrates how the spatial profiles of modes $124$ and $129$ change across the avoided crossing. Initially, at $\alpha = 0$, the blue and red modes have distinct profiles and correspond to points 1 and 4, respectively. Near the avoided crossing (points 2 and 5), both states undergo significant deformation and their amplitude distributions  $|\phi_{124}^{(\balpha)}|$ and $|\phi_{129}^{(\balpha)}|$ become similar. After the avoided crossing (points 3 and 6), the modes from blue and red branches resemble the red and blue modes at $\alpha = 0$, respectively, demonstrating an ``exchange'' of spatial structure. This behavior is further confirmed in Fig.~\ref{fig2}(c), which plots the center-of-mass coordinates ${\bm r}_{i}^{(\alpha)}=\left( x_c,y_c \right)$ of the two modes as functions of $\alpha$, showing a smooth exchange of position consistent with the avoided crossing scenario.

A more intricate transformation of the states is observed for branches $127$ (yellow), $135$ (magenta), and $137$ (green). As shown in Fig.~\ref{fig2}(a), the magenta branch undergoes two avoided crossings: the first near $\alpha \approx 0.001$ with the green branch, and the second near $\alpha \approx 0.005$ with the yellow one. This suggests that during deformation in the parameter space, resonant interactions with different modes may occur at different values of $\alpha$. The corresponding trajectories of the centers of mass are presented in Fig.~\ref{fig2}(d), also revealing a more complex behavior: mode $135$ first exchanges its position with mode $137$, and later with mode $127$ for larger $\alpha$. This is consistent with the behavior of the respective propagation constants.

One of the key factors in the 2D case is the direction of the gradient determined by angle $\gamma$, as it can dramatically alter propagation dynamics. This sensitivity of the modes to the gradient direction is demonstrated in Fig.~\ref{fig2}(b), showing dependence of eigenvalues $\beta_i^{(\balpha)}$ on $\alpha$ for angle $\gamma = -\pi/5$. In comparison with the previous case, the change in gradient direction leads to a complete restructuring of the branches in the spectrum for $\alpha > 0$. For instance, the blue branch, which previously had a positive slope, now exhibits a negative slope and undergoes two avoided crossings with branches originating from states $116$ and $129$. These interactions are absent for the previous angle $\gamma = \pi/3$. As we will show below, this change in spectral structure has direct consequences for the dynamics: varying the gradient direction can completely change the set of resonantly interacting modes, even when the initial condition remains the same.

If a given localized mode, obtained for $\alpha=0$, is used at the input in the lattice with a nonzero, but small gradient $\alpha$, it may, in practice, excite only a few localized modes of the lattice. Thus, to address the central question of this work, one has to determine the modes of $H_{\bm \alpha}$ that can be efficiently excited by such input. Clearly, due to localization, any newly excited mode should have its center of mass sufficiently close to that of the input mode (see more details below). This observation represents an additional justification for restricting the numerical analysis to the effective area, which in Fig.~\ref{fig1} and Fig.~\ref{fig2} (e) is delimited by the circumference of the radius $\rho=30$. All modes with appreciable overlap that can participate in the dynamics are confined to the effective area. It is worth noting that the centers of mass of all extended modes, which are located near the origin, also lie within this region. However, such modes are not excited in a system with a weak gradient due to a relatively large difference between their propagation constants and propagation constants of the selected localized states and, therefore, will not be considered further.

Now we introduce two bases $\bphi^{(0)}$ and $\bphi^{(\balpha)}$:
\begin{eqnarray*}
	\bphi^{(0)}=(\phi_1^{(0)}, ...,\phi_N^{(0)})^T
	\,\, \mbox{and}\,\,
	\bphi^{(\balpha)}=(\phi_1^{(\balpha)}, ...,\phi_N^{(\balpha)})^T
\end{eqnarray*}
($T$ stands for transpose) which are considered complete in the subspace of localized modes in the effective area. These bases are connected by the unitary transformation
\begin{eqnarray}
\label{unitary}
	\bphi^{(\balpha)}=\bm{S}_{\balpha} \bphi^{(0)}, \qquad S_{ij}^{(\balpha)} =\langle\phi_{j}^{(0)},\phi_i^{(\balpha)}\rangle.
\end{eqnarray}
Respectively, one can expand 
\begin{eqnarray}
\label{expansion}
	\Psi=\sum_{i=1}^{N}c_i^{(0)}(z)\phi_{i}^{(0)}({\bm r})
	=\sum_{i=1}^{N}c_i^{(\balpha)}(z)\phi_{i}^{(\balpha)}({\bm r})
\end{eqnarray}
and introduce the center of mass of the field: 
\begin{eqnarray}
\label{bR}
	{\bm R}=\langle \Psi,{\bm r}\Psi\rangle ={\bm{c}}_0^\dagger{\bm R}^{(0)}{\bm{c}}_0={\bm{c}}_{\balpha}^\dagger{\bm R}^{(\balpha)}{\bm{c}}_{\balpha},
\end{eqnarray}
where ${\bm R}^{(\balpha)}$ is the matrix with (vector) elements
\begin{eqnarray}
\label{r_elem}
	{\bm r}_{ji}^{(\balpha)}=\langle\phi_{j}^{(\balpha)},{\bm r} \phi_i^{(\balpha)}\rangle, 
\end{eqnarray}    
 and
${\bm{c}}_{\balpha}=(c_1^{(\balpha)}, ...,c_N^{(\balpha)})^T$. It follows from (\ref{unitary}) that
\begin{eqnarray}
\label{cSc}
    {\bm{c}}^{(0)}= \bm{S}_{\balpha} {\bm{c}}^{(\balpha)}.
\end{eqnarray}

The evolution of the coefficients is determined by the equations 
\begin{eqnarray}
	\label{dc_0}
	i\frac{d{\bm{c}}_{0}}{dz}=&-({\bm{B}}_{0}+\balpha\cdot{\bm R}^{(0)}){\bm{c}}_{0}
\end{eqnarray}
in the original basis and 
\begin{eqnarray}
	\label{dc_al}
	i\frac{d{\bm{c}}_{\balpha}}{dz}=&-{\bm{B}}_{\balpha}{\bm{c}}_{\balpha}
\end{eqnarray}
in the basis $\phi_i^{(\balpha)}$, where $ {\bm{B}}_{\balpha}=\mbox{diag}(\beta_1^{(\balpha)},..., \beta_N^{(\balpha)}) $. Equation (\ref{dc_al}) is readily solved.
Finally, combining (\ref{bR}) and (\ref{cSc}) with the solution of (\ref{dc_al}), one obtains the formula for the evolution of the center of mass:
\begin{eqnarray}
	\label{COM}
	{\bm R}(z)={\bm{c}}_{0}^\dagger(0) \bm{S}_{\balpha}^\dagger \bLambda_{\balpha}^\dagger(z)\bm{S}_{\balpha}{\bm R}^{(0)}
\bm{S}_{\balpha}^\dagger\bLambda_{\balpha}(z)\bm{S}_{\balpha}{\bm{c}}_{0}(0),
\end{eqnarray}
where
$
	\bLambda_{\balpha}(z)=\mbox{diag}(e^{i\beta_1^{(\balpha)}z},...,e^{i\beta_N^{(\balpha)}z}).
$	
We emphasize that the only approximation made so far was the absence of excited modes below the ME and beyond the effective area.  

\subsection{Selection rule}

\begin{figure}[t]
\centering
\includegraphics[width=1.00\linewidth]{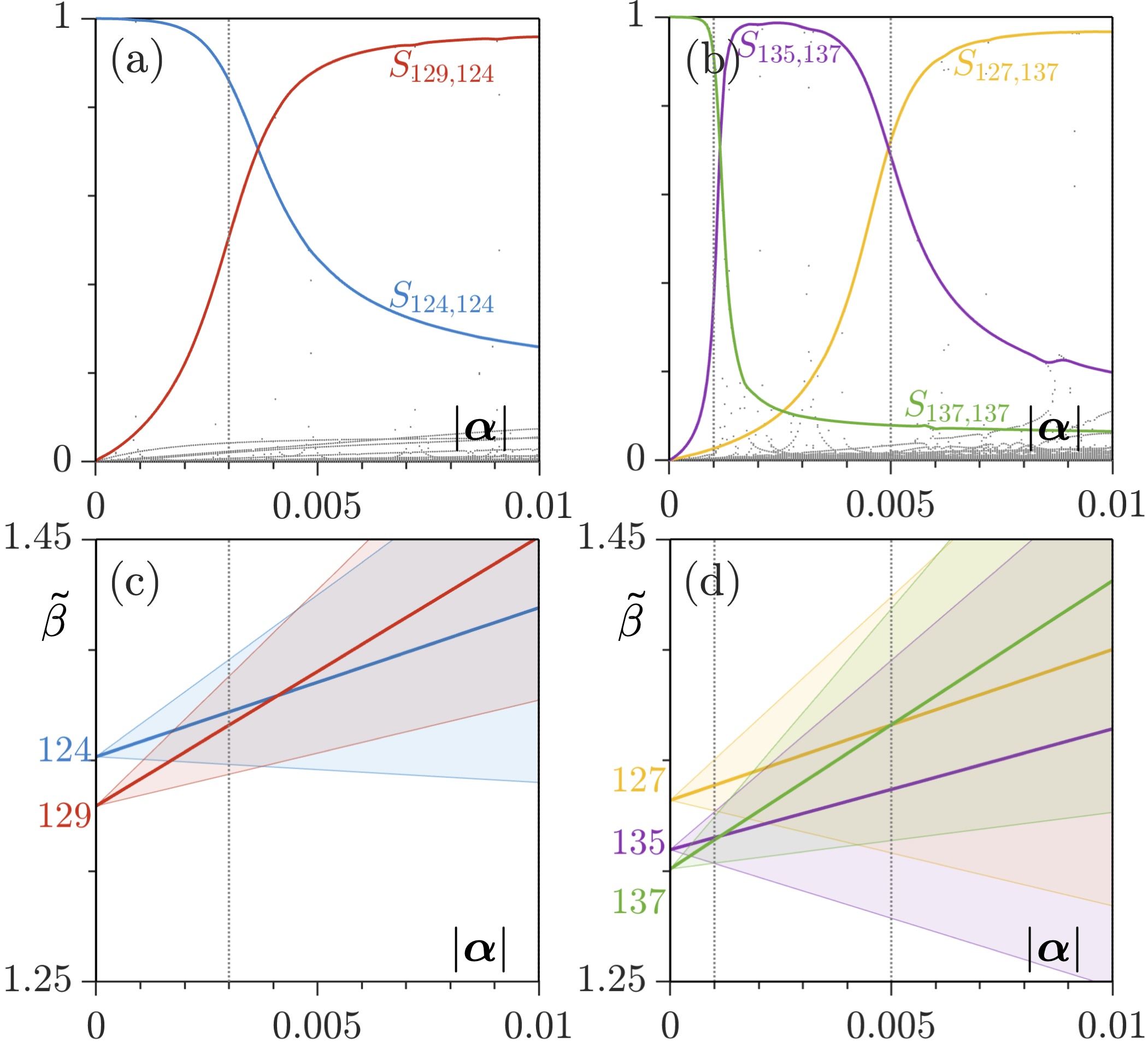}
\caption{Variation of the overlap integrals $S_{ij}^{(\balpha)}$ with $|\balpha|$ for   ${j}=124$ (a) and ${j}=137$ (b). Centers ${\tilde\beta}_i^{(\balpha)}$ (solid lines) of the Gershgorin intervals (shaded regions) as functions of $|\balpha|$ for the modes $124$, $129$ (c) and $127$, $135$, and $137$ (d). In all panels $\gamma=\pi/3$. Dashed vertical lines indicate the values of $|\balpha|$ corresponding to the dynamic examples shown in Figs.~\ref{fig4} and~\ref{fig5}. }
\label{fig3}
\end{figure}

Generally, a given input beam can excite multiple localized modes. Thus, to pursue the main objective outlined above, we should first address the following auxiliary task: {\em given an initial state that is an eigenmode of the Hamiltonian $H_0$ without a gradient ($\alpha=0$), determine which modes will be significantly excited for $0<\alpha\ll 1$ at $z>0$.}
This task is addressed below in this subsection.

The initial condition that corresponds to excitation of $j$'s mode of $H_0$ is $c_i^{(0)}(0)=\delta_{ij}$. Respectively, the amplitude of the state $\phi_{i}^{(\balpha)}(0)$ at $z=0$ is determined by $c_i^{(\balpha)}(0)=S_{ij}^{(\balpha)}$. The modes for which $c_i^{(\balpha)}(0)$ is negligible are regarded as unexcited. To quantify this, we introduce a parameter $0<\Delta\ll 1$ and consider only the modes $j$ with non-negligible projections
\begin{eqnarray}
	\label{position}
    |S_{ij}^{(\balpha)}| >\Delta.
\end{eqnarray}
Following~\cite{Prates2024}, we call this requirement as {\em spatial proximity conditions}. In numerical simulations shown below for identification of the interacting modes, we set $\Delta= 0.1$. Only the modes satisfying (\ref{position}) can effectively participate in the energy exchange with the input mode $\phi_j^{(0)}$, while all other modes, even being excited, have negligible weights and do not affect the dynamics. In Fig.~\ref{fig3}(a) and (b), we illustrate the dependence of elements $S_{ij}^{(\balpha)}$ for all modes $\phi_i^{(\balpha)}$ from the effective area on gradient $\alpha$ for the two input modes $j=124$ (panel (a)) and $j=137$ (panel (b)). In panel (a), one observes that $S_{124,124}^{(\balpha)}$ (blue line) remains close to 1 for small $\alpha$, then decreases monotonically as $\alpha$ increases. In contrast, $S_{129,124}^{(\balpha)}$ (red line) is negligible at $\bm \alpha = 0$, but rises significantly, reaching a value close to $1$ at $\alpha = 0.01$. The two curves intersect and take on similar values around the avoided crossing in the spectrum at approximately $\alpha \approx 0.003$, indicating strong modal overlap. This behavior reflects how the evolving modes for different $\alpha$ overlap with the initial mode $\phi_{124}^{(0)}$. At $\alpha = 0$, the centers of mass of modes $\phi_{124}^{(0)}$ and $\phi_{129}^{(0)}$ are spatially separated, as shown in Fig.~\ref{fig2}(c) and (e) at points 1 and 4. As $\alpha$ increases, the mode profiles deform and their centers of mass approach, as seen at points 2 and 5, leading to comparable values of $S_{i,124}^{(\balpha)}$ for $i=124$ and $i=129$. With further increase in $\alpha$, the centers of mass separate again (points 3 and 6), and only $\phi_{129}^{(\balpha)}$ retains significant overlap with the initial mode. For all other localized modes (shown in gray), we obtain $|S_{{ i,124}}^{(\balpha)}|<0.1$ for all $\alpha<0.01$ in Fig.~\ref{fig3}(a) and $|S_{{ i,137}}^{(\balpha)}|<0.1$ for almost all $\alpha<0.01$ (except for extremely short intervals, which are omitted from our consideration due to their narrowness) in Fig.~\ref{fig3}(b). We also note that in Fig.~\ref{fig3}(a) at $\alpha\gtrsim 0.005$ the mode 129 has a larger overlap integral with input mode 124, than mode 124 with the same index.

Besides the elements of the overlap matrix $\bm{S}_{\balpha}$, the evolution of the center of mass ${\bm R}(z)$ of the wave\-pack\-et along $z$ axis, is determined by the exponential factors $\exp[i(\beta_{i}^{(\balpha)}-\beta_{j}^{(\balpha)})z]$ stemming from the $\bLambda_{\balpha}(z)$ matrices in expression (\ref{COM}).  Thus, in addition to (\ref{position}), effectively interacting modes must obey sufficiently close propagation constants, i.e., for $i\neq j$, one  requires
\begin{eqnarray}
\label{reson1}
    |\beta_{i}^{(\balpha)}-\beta_{j}^{(\balpha)}|\ll |\beta_{i}^{(\balpha)}|, |\beta_{j}^{(\balpha)}|.
\end{eqnarray}
To express this condition through the characteristics of the modes of the system without gradient (i.e., at $\alpha=0$, as it is required by the formulated setting of the problem) we recall that the bases are related by the unitary transformation (\ref{unitary}) and hence, the $\beta_{i}^{(\balpha)}$ are the eigenvalues also of the matrix $-({\bm{B}}^{(0)}+\balpha\cdot{\bm R}^{(0)})$ whose diagonal elements are denoted by  
\begin{eqnarray}
\tbeta_i^{(\balpha)}=\beta_i^{(0)}+\balpha\cdot{\bm r}_{ii}^{(0)}.
\end{eqnarray}
For several modes (explored below) the dependencies $\tbeta_i^{(\balpha)}$ {\it versus} $\alpha$ are illustrated by the solid lines in Figs.~\ref{fig3}(c) and (d).

Importantly, $\tbeta_i^{(\balpha)}$ are determined by the eigenvalues of the untilted lattice and by the gradient. Now we must relate them to the propagation constants $\beta_{j}^{(\balpha)}$ determining  (quasi-)resonances (\ref{reson1}). To this end, we recall that according to the Gershgorin circle theorem~\cite{Gershgorin} for each $\beta_{i}^{(\balpha)}$ one can find $\tbeta_{j}^{(\balpha)}$ such that 
\begin{eqnarray}
	\label{energy0}
	|\beta_{i}^{(\balpha)}-\tbeta_{j}^{(\balpha)}|\leq \rho_j^{(\balpha)} ,
\end{eqnarray}
where
\begin{eqnarray}
    \label{GR}
    \rho_j^{(\balpha)}=\sum_{j'=1\atop j'\neq j}^{N}|\balpha\cdot {\bm r}_{jj'}^{(0)}|
\end{eqnarray}
are the Gershgorin radii. Since we consider real spectra, as a matter of fact, $\rho_j^{(\balpha)}$ are half-lengths of the intervals. In Figs.~\ref{fig3}(c) and (d), they are illustrated by shaded regions around the respective  $\tbeta_{j}^{(\balpha)}$ lines. Note that generally speaking, $i$ can be different from $j$. Thus, if we require that the respective Gershgorin intervals overlap for a given gradient $\balpha$ and, i.e., if 
\begin{eqnarray}
	\label{energy}
	|\tbeta_{j}^{(\balpha)}-\tbeta_{j'}^{(\balpha)}|\leq \rho_j^{(\balpha)}+\rho_{j'}^{(\balpha)}
\end{eqnarray}
for the modes $i$ and $i'$ of $H_{\balpha}$, then one can find the respective modes $j$ and $j'$ of $H_0$ such that 
\begin{eqnarray}
\label{resonance2}
      |\beta_{i}^{(\balpha)}-\beta_{i'}^{(\balpha)}|\leq 2 (\rho_{j}^{(\balpha)} +\rho_{j'}^{(\balpha)})  .
\end{eqnarray}
Thus, for the gradients small enough, one can always find resonant modes, satisfying (\ref{reson1}). 

Strictly speaking, the derived conditions do not solve the formulated task yet, because having considered modes of $H_0$ with indices $j$ and $j'$, we have not determined yet the modes $\phi_i^{(0)}$ and $\phi_{i'}^{(0)}$, which will interact resonantly: formally they can be different from the modes $\phi_j^{(0)}$ and $\phi_{j'}^{(0)}$, which give origin to Gershgorin intervals. To complete this last step, we recall that in the limit $\alpha\to 0$ the Gershgorin intervals collapse to points $\beta_i^{(0)}$ and remain non-overlapped for sufficiently small $\alpha$ [see examples in Figs.~\ref{fig3}(c) and (d)]. Due to continuous dependence of $\beta_i^{(\balpha)}$ on $\balpha$, it follows that $\beta_i^{(\balpha)}$ belongs to the Gershgorin interval originated at $\beta_i^{(0)}$ even when the intervals start to overlap. Thus, for sufficiently small $\alpha$, one has $i=j$ and $i=j'$. In Fig.~\ref{fig3}(c) and (d), we illustrate quasi-resonances and Gershgorin intervals for several modes.

The condition (\ref{resonance2}) invoking the input modes of the Hamiltonian $H_0$, can be viewed as a {\em quasi-res\-o\-nance} in the ``energy'' space, representing the proximity of the propagation constants. Conditions (\ref{position}) and (\ref{resonance2}) verified simultaneously constitute the {\em selection rule} determining the modes of $H_0$ in the coordinate and energy spaces, that may interact resonantly with a given input mode. In other words, these conditions determine modes that can be involved in the energy exchange upon propagation. Each of those conditions alone is not sufficient for efficient coupling between modes.

Finally, we note that although in the 2D case the exact resonance at the level crossing $\beta_{i}^{(\balpha)}=\beta_{i'}^{(\balpha)}$ is possible [examples can be found in Figs.~\ref{fig1}(a) and (b)], in a generic situation it will not result in resonant interactions of the modes, because of negligible hopping integrals $\langle\phi_j^{(0)},\phi_{j'}^{(\alpha)}\rangle$. In a general situation, the quasi-resonance occurs at avoided crossings [like the ones shown in Fig.~\ref{fig1} (a) and (b)].

\begin{figure}[t]
\centering
\includegraphics[width=1.00\linewidth]{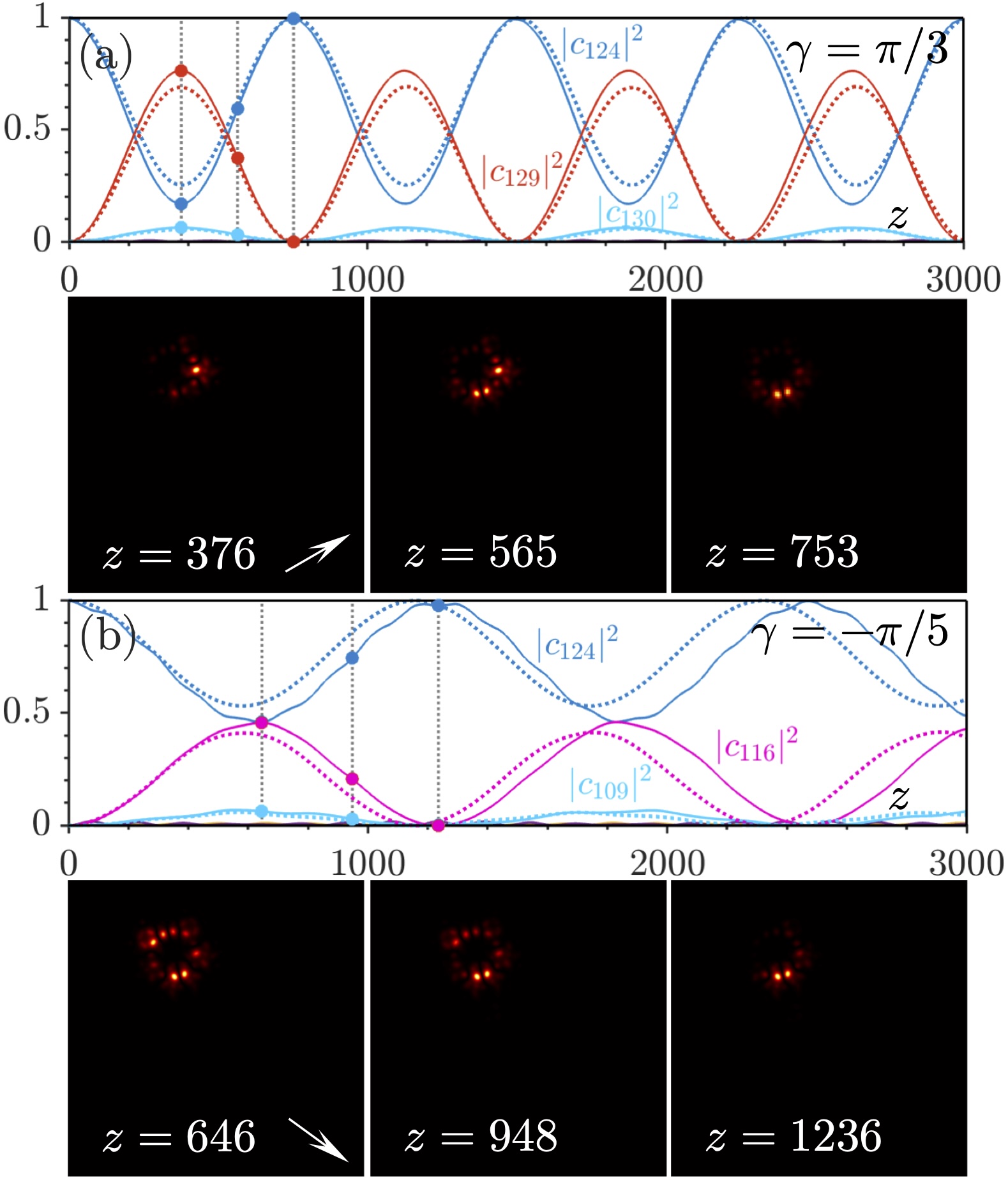}
\caption{Evolution of weights $|c_i |^2$ of all linear modes of the system at $|\balpha|=0$ (top) and field modulus distributions at different values of $z$ corresponding to the dots and vertical dashed lines in the top panels (bottom) for the angle $\gamma=\pi/3$ (a) and $\gamma=-\pi/5$ (b) for $|\balpha|=0.003$. The initial state is mode $n=124$ of the system with $|\balpha|=0$. Arrows indicate the direction of the gradient. The dashed lines show the analytical solution.}
\label{fig4}
\end{figure}

\subsection{Examples of BLZ oscillations}

\subsubsection{Two-mode dynamics}

Now we turn to the examples of different propagation regimes and BLZ oscillations. For the illustration purposes we choose the input mode $\phi_{124}^{(0)}$ (the mode labeled as point 1 in Fig.~\ref{fig2}) for the lattice with $\gamma=\pi/3$. At sufficiently small $\alpha\lesssim 0.001$ one cannot encounter other modes satisfying both the proximity condition (\ref{position}) [Fig.~\ref{fig3}(a)] and quasi-resonance condition (\ref{resonance2}) [Fig.~\ref{fig3}(c)]. For this region of parameters, in spite of the change of the propagation constant (blue line in Fig.~\ref{fig2}) due to the gradient, the mode exhibits no significant dynamics beyond weak amplitude fluctuations (not shown here). With further increase of gradient $\alpha$ both spatial and quasi-resonant conditions for efficient arise between modes 124 and 129, as evidenced by the avoided level crossing, see modes corresponding to points 2 and 5 in Fig.~\ref{fig2}(a) indicated by the vertical dashed line in Fig.~\ref{fig3}(a) and (c). Now, large amplitude (in the sense of amount of the transferred power) BLZ oscillations involving these two modes occur. The results of the direct numerical simulations of these oscillations are shown in Fig.~\ref{fig4} (a). The shown oscillations persist over long propagation distances. This extreme robustness of the BLZ oscillations arises from suppressed diffraction in the incommensurate moir\'e lattices (observed earlier in different experimental settings~\cite{Wang2020,Fu2020}). The period of the oscillations in theory is determined by the formula $Z_{ij}=2\pi/|\beta_i^{(\balpha)}-\beta_j^{(\balpha)}|$. Estimating this quantity for $\beta_{124}^{(\balpha)}$ and $\beta_{129}^{(\balpha)}$ at $\alpha=0.003$ (see Fig.~\ref{fig2}(a)) one obtains $Z\approx 761.8$ what matches well the period observed in direct simulations [Fig.~\ref{fig4} (a)]. In the dynamical figures, the prediction of the analytical theory is shown as dashed lines and closely matches numerical results.

A remarkable feature of BLZ oscillations in moir\'e lattices is that increasing the gradient does not necessarily intensify power exchange between modes; on the contrary, it can have the opposite effect, weakening the coupling and thereby damping the beam’s oscillations. Indeed, our numerical simulations (not shown here) confirm that a further increase of the gradient results in much weaker power exchange between modes 124 and 129. Note that at $\alpha=0.01$ the Gershgorin radii, although still overlapping [Fig.~\ref{fig3} (c)], are relatively large such that the quasi-resonant condition~(\ref{reson1}) is no longer satisfied. In this regime, the only noticeable impact of increasing the gradient is a higher frequency of the low‑amplitude oscillations of the mode intensity, instead of the enhancement of power transfer.

As it was mentioned above, the direction of the gradient plays a crucial role. Fig.~\ref{fig4}(b) shows the dynamics for the same initial state $\phi_{124}^{(0)}$ in a system with the same gradient $\alpha = 0.003$, but with angle $\gamma = -\pi/5$. As evident from the figure, the energy is now transferred to a completely different mode, specifically, to the one labeled 116, demonstrating the strong dependence of the dynamics on the gradient direction. Note that while all modes within the effective area were accounted, in the upper panels of Fig.~\ref{fig4}(a) and Fig.~\ref{fig4}(b) one can distinguish contributions from only three modes (with two modes dominating the dynamics in each case): the input mode 124 and resonantly interacting mode 129 at $\gamma=\pi/3$ [panel (a)] and 109 at $\gamma=-\pi/5$ [panel (b)]. This reveals the effect of the direction of the gradient on the BLZ oscillations, which is shown by the arrows in the leftmost snapshots of the dynamics. Remarkably, one observes that the positions of the modes in resonance do not have any visual connection with the direction of the gradient, which is in full agreement with the theoretical predictions since the selection rules do not depend on the angle $\gamma$ in a direct way. 
It is important to note that changing the orientation of the gradient $\gamma$ can not only alter the resonant state involved in power exchange, but in certain directions, it can completely suppress this exchange, causing the initial state to propagate with minimal oscillations. For example, at a fixed $\alpha=0.003$, the gradient angle can be tuned such that only weak amplitude oscillations occur, without any significant mode coupling.

\begin{figure}[t!]
\centering
\includegraphics[width=1.00\linewidth]{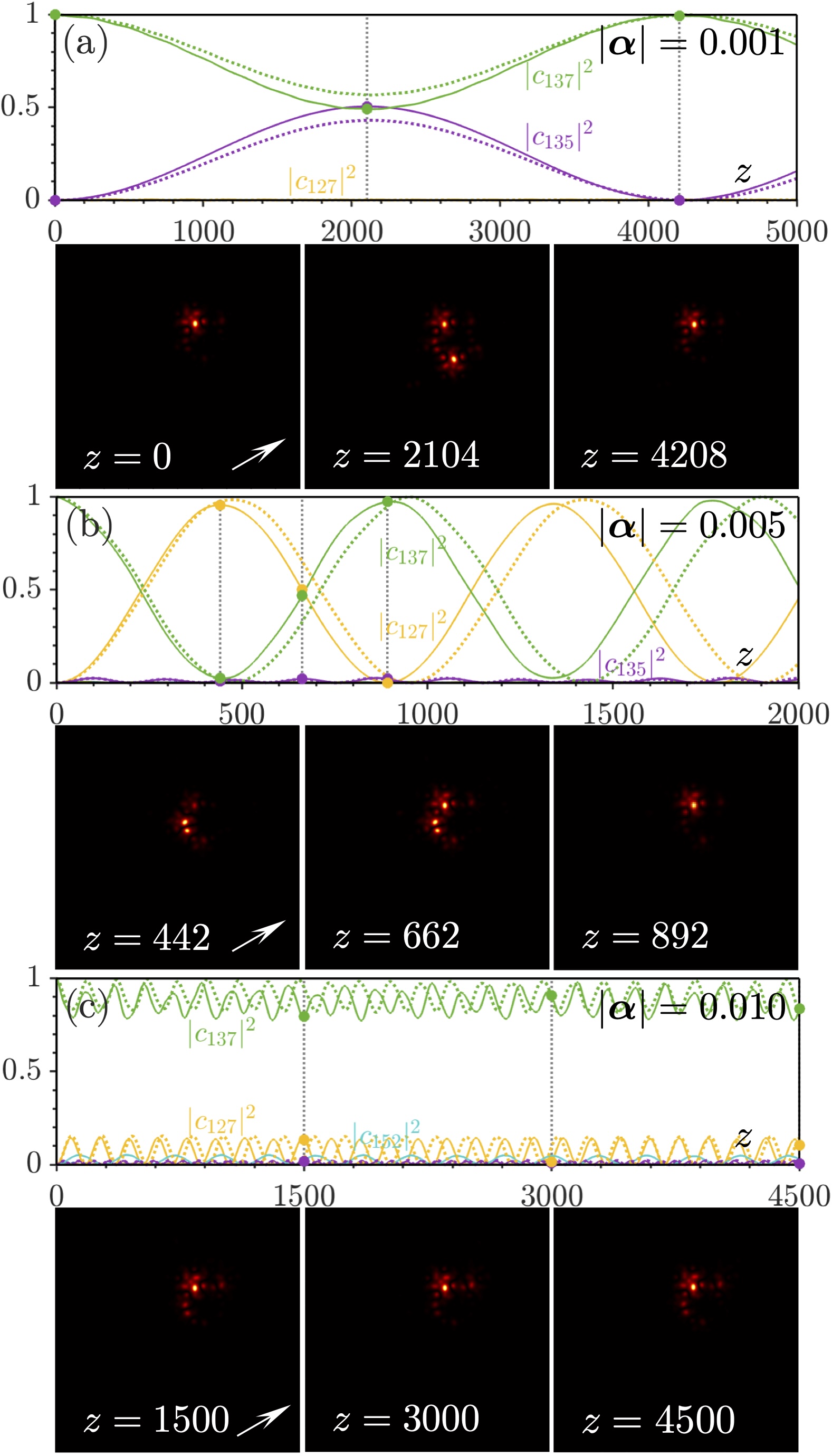}
\caption{Evolution of weights $|c_n |^2$ of all linear modes of the system at $|\balpha|=0$ for the angle $\gamma=\pi/3$ (top) and field modulus distributions at different values of $z$ corresponding to the dots and vertical dashed lines in the top panels (bottom) for $|\balpha|=0.001$ (a), $|\balpha|=0.005$ (b), and $|\balpha|=0.010$ (c). The initial state is mode $n=137$ of the system with $|\balpha|=0$. Arrows indicate the direction of the gradient. The dashed lines show the analytical solution.}
\label{fig5}
\end{figure}

\subsubsection{Oscillations involving several modes}

An even more intricate dependence of the dynamics on the gradient $\alpha$ emerges when different modes satisfy the selection rule at different values of $\alpha$. To illustrate this, consider the input mode $\phi_{137}^{(0)}$ propagating in the lattice with angle of gradient $\gamma=\pi/3$. The change of its propagation constant upon change of the gradient is shown by the green curve in Fig.~\ref{fig2}(a), while the corresponding coupling coefficients $S_{i,137}^{(\balpha)}$ and Gershgorin intervals are presented in Fig.~\ref{fig3}(b) and (d), respectively. As these plots indicate, the selection rule is fulfilled for mode 135 around $\alpha \approx 0.001$. With increasing $\alpha$, this resonance condition is no longer met for mode 135 but becomes valid for mode 127 near $\alpha \approx 0.005$. Beyond this, up to $\alpha = 0.01$, no modes satisfy the selection criteria. This behavior is confirmed by direct propagation simulations of mode $\phi_{137}^{(0)}$ under different gradients. At $\alpha = 0.001$, energy transfer between modes 137 and 135 is observed [Fig.~\ref{fig5}(a)]. As the gradient increases to $\alpha = 0.005$, the interaction with mode 135 is suppressed, and coupling with mode 127 appears [Fig.~\ref{fig5}(b)]. Once again, increasing the gradient does not necessarily enhance mode coupling; in fact, for $\alpha = 0.01$, energy exchange is almost completely suppressed. This is evident from the snapshots shown in Fig.~\ref{fig5}(c), where the amplitude profile remains nearly unchanged during the propagation.

\begin{figure}[t!]
\centering
\includegraphics[width=1.00\linewidth]{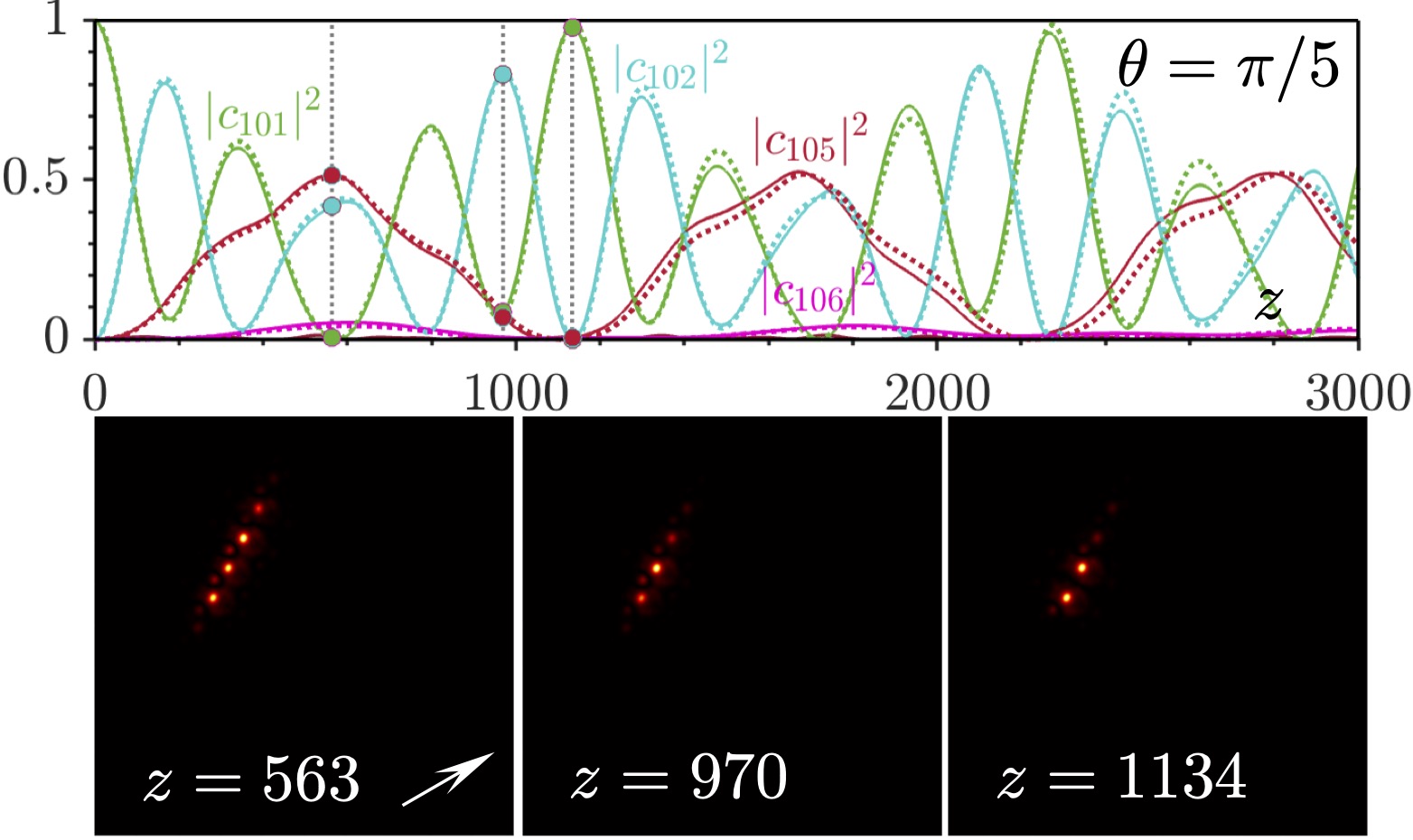}
\caption{{Evolution of weights $|c_n |^2$ of all linear modes of the system at $|\balpha|=0$ for the angle between sublattices $\theta=\pi/5$, gradient orientation $\gamma=\pi/3$, and $|\balpha|=0.005$ (top) and field modulus distributions at different values of $z$ corresponding to the dots and vertical dashed lines in the top panels (bottom). The initial state is mode $n=101$ of the system with $|\balpha|=0$. The arrow indicates the direction of the gradient. The dashed lines show the analytical solution.}}
\label{fig6}
\end{figure}

\subsubsection{{Multimode oscillations for a different sublattice orientation}}

{{ The described physics of BLZ oscillations is confirmed by numerical simulations performed for other sublattice orientations. Figure~\ref{fig6} illustrates the evolution of the mode weights $|c_n|^2$ of all linear modes at $|\balpha| = 0$ for the system with $\gamma = \pi/3$ and $|\balpha| = 0.003$, obtained for the angle between sublattices $\theta = \pi/5$. The initial state is the mode $n = 101$. For this mode, the spatial proximity condition $|S_{i,101}^{(\balpha)}| > \Delta$ is satisfied only for $i = 102$, $105$, and $106$, while overlap with other modes is negligibly small. For these three states, the spectral proximity condition is fulfilled as well. Consistently, Fig.~\ref{fig6} demonstrates pronounced power exchange among modes $101$, $102$, and $105$, with weaker coupling to mode $106$, whereas interactions with all other modes remain suppressed. The lower panels of Fig.~\ref{fig6} display the field modulus distributions at different propagation distances $z$, marked by the vertical dashed lines in the top panel. These snapshots indicate that the power transfer occurs predominantly along a single spatial line, which is noticeably misaligned with the gradient direction $\gamma$. Finally, the analytical predictions for weights $|c_n|^2$ dynamics (dashed curves) are in excellent agreement with the numerical simulations.
}}

\begin{figure}[t!]
\centering
\includegraphics[width=1.00\linewidth]{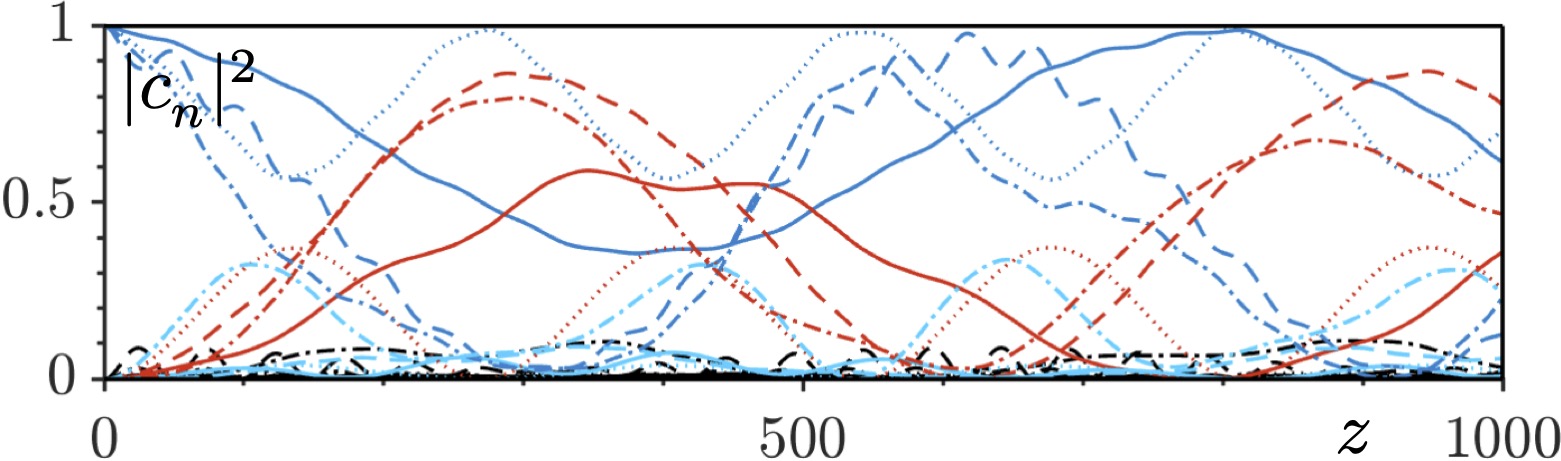}
\caption{{Evolution in the presence of disorder with $\delta_p = 0.02$ and $\delta_r = 0.01$ of the mode weights $|c_n|^2$ of all linear modes of the unperturbed system at $|\balpha| = 0$, for gradient orientation $\gamma = \pi/3$ and gradient strength $|\balpha| = 0.003$. The initial state corresponds to mode $n = 124$ of the unperturbed ($|\balpha|=0$) system. Different line styles represent distinct disorder realizations. The arrow indicates the gradient direction. Dashed lines represent the analytical predictions.}}
\label{fig7}
\end{figure}

\subsubsection{{Oscillatory dynamics in a perturbed system}}
\label{subsec:perturbation}

{{
To examine the structural robustness of the BLZ dynamics we simulated propagation of a beam in a moir\'e lattice where the waveguide depths of the sublattices were randomly distributed within the interval $[p_j(1-\delta_p), p_j(1+\delta_p)]$, $j=1,2$, and the waveguide positions in both the $x$ and $y$ directions were shifted by random values in the range $[-\delta_r, \delta_r]$. We used parameters from Fig.~\ref{fig4}(a) (i.e., $\gamma = \pi/3$, $|\balpha| = 0.003$, and $\theta = \pi/3$) simulated now for $\delta_p = 0.02$ and $\delta_r = 0.01$. We used the same input mode, $n = 124$, of the unperturbed lattice without gradient, $|\balpha|=0$ as in Fig.~\ref{fig4}(a). 
In Fig.~\ref{fig7} we show the dynamics of the beam projections $|c_i|^2$ 
on the eigenmodes of the unperturbed lattice, with different line styles representing distinct disorder realizations.
Although the oscillation period and amplitude are moderately modified compared to the unperturbed case, the mode composition remains unchanged, and oscillations persist even when the excitation is not an eigenmode of the perturbed system. 
We emphasize that for each realization of disorder the eigenmodes of the perturbed system can be determined, by analogy with the unperturbed moir\'e lattice, and the spectrum remains discrete. Thus, our theoretical framework can then be used to predict the corresponding mode composition and fully describe the BLZ dynamics. The selection rule still determines the set of excited modes, however, a thorough analytical description requires a full numerical analysis of eigenmodes of the randomized lattice. It should be mentioned that for $|\balpha| = 0.003$, the dominant modes exhibit dynamics comparable to those of the unperturbed system, whereas for larger values of $|\balpha|$, the effect of disorder becomes more pronounced, occasionally inducing power transfer to additional modes not involved in the unperturbed dynamics.
}}

\begin{figure*}[t!]
\centering
\includegraphics[width=1.00\linewidth]{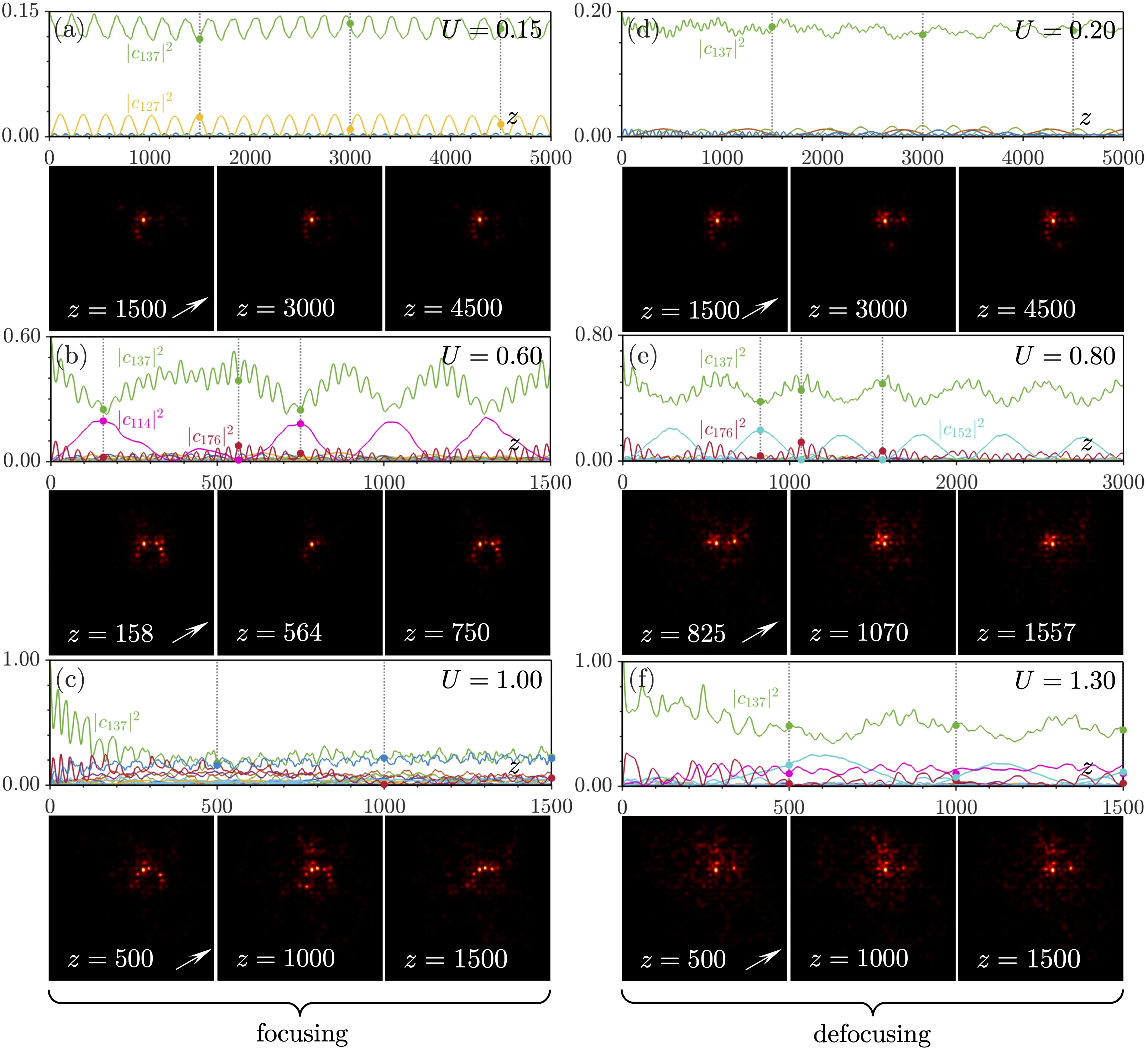}
\caption{Evolution of weights $|c_n |^2$ of all linear modes of the system at $|\balpha|=0$ for the angle $\gamma=\pi/3$ (top) and field modulus distributions at different values of $z$ corresponding to the dots and vertical dashed lines in the top panels (bottom). Results are shown for nonlinear propagation in focusing (a–c) and defocusing (d–f) media with $|\balpha|=0.005$, for input powers: $U=0.15$ (a), $U=0.60$ (b), $U=1.00$ (c), $U=0.20$ (d), $U=0.80$ (e), and $U=1.30$ (f). The initial state is mode $n=137$ of the system with $|\balpha|=0$. Arrows indicate the direction of the gradient. The dashed lines show the analytical solution.}
\label{fig8}
\end{figure*}
 
\section{Nonlinear BLZ oscillations}

 Now we turn to the effect of weak nonlinearity of the BLZ oscillations governed by the model
 \begin{eqnarray}
	\label{main-nonlin}
	i\frac{\partial \Psi}{\partial z}=H_{\balpha}\Psi+g|\Psi|^2\Psi ,
\end{eqnarray}
where $g=1$ and $g=-1$ describe defocusing and focusing nonlinearities, respectively, while $H_{\balpha}$ is the Hamiltonian introduced in (\ref{Hamilt}). Since our consideration is limited to the localized modes, neglecting excitation of extended states, in the nonlinear case, one can still search for a solution of (\ref{main-nonlin}) in the form of the expansion over the basis $\phi_i^{(0)}$ [see (\ref{expansion})]. For sufficiently small input intensities, where only localized modes are excited at $z=0$, this condition can be expressed as $U=\sum_{j=1}^N|c_j|^2\ll 1$. In the leading order, substituting the ansatz into Eq.~(\ref{main-nonlin}) yields the following system for the modal amplitudes:
\begin{eqnarray}
	\label{dc_0_nonlin}
	i\frac{d{c}_j}{dz}=-\beta_j^{(0)}c_j-\sum_{k=1}^N(\balpha\cdot{\bm R}^{(0)})_{jk}{c}_k+\chi_j|{c}_j|^2{c}_j ,
\end{eqnarray}
where $\chi_j=g\langle[\phi_j^{(0)}]^2,[\phi_j^{(0)}]^2\rangle$ is the effective nonlinearity (to shorten the notations we do not use the upper index $0$ in the coefficients $c_j(z)$, since only the basis of $H_0$ is considered below). In the system (\ref{dc_0_nonlin}) we neglected all the terms proportional to  $\langle \phi_{i_1}^{(0)}\phi_{i_2}^{(0)},\phi_{i_3}^{(0)},\phi_{i_4}^{(0)}\rangle$, except the terms where all coefficients are equal $i_1=\cdots=i_4$, i.e., $\chi_j$. This is justified by the fact that such overlap integrals are significant only for modes that are spatially close to each other (see, e.g., the 1D numerical examples in~\cite{Konotop2024}). Meanwhile, due to the selection rule described above, the modes resonantly interacting at nonzero gradient are typically located relatively far apart at $\alpha=0$ [see Fig.~\ref{fig1}(b), Fig.~\ref{fig2}(d), and Fig.~\ref{fig3}(a),(b)]. We can thus conclude that the nonlinear term in Eq.~(\ref{main-nonlin}) is expressed predominantly through the self‑action of the modes. The model describes the dynamics well for small input powers, specified above (not shown here).

As an example of light propagation in a nonlinear medium with a refractive index gradient at larger nonlinearities, we consider the initial mode $\phi_{137}^{(0)}$ in a system with $\alpha = 0.005$ and $\gamma = \pi/3$. The same configuration was used for the linear dynamics in Fig.~\ref{fig5}(b). Nonlinear evolution for different input powers $U=|c_j|^2$ is shown in Fig.~\ref{fig8}, with the left column [Figs.~\ref{fig8}~(a)–(c)] corresponding to a focusing medium and the right column [Figs.~\ref{fig8}~(d)–(f)] to a defocusing one. In the linear case, power transfer occurred between modes 137 and 127. However, under weak nonlinearity (e.g., $U=0.15$ for focusing, $U=0.20$ for defocusing), this transfer is suppressed, as evidenced in Figs.~\ref{fig8}(a) and (d): only minor variations in modal weights occur, and the field profiles remain nearly unchanged. Notably, model (\ref{dc_0_nonlin}) predicts suppression of oscillations for these modes 137 and 127 as power increases; however, at this relatively high power level, it no longer accurately captures the amplitude and period of these low-amplitude dynamics. For lower initial powers $U$ (not shown here), the predictions of this model remain reasonably accurate. As the input power increases, nonlinear effects can bring new modes into resonance. For instance, at $U=0.60$ in the focusing regime [Fig.~\ref{fig8}(b)], coupling emerges primarily with mode 114, and to a lesser extent with other modes, resulting in irregular dynamics. In the defocusing case at the same power [Fig.~\ref{fig8}(e)], modes 152 and 176 are excited instead, both having lower propagation constants than the initial mode.
At even higher input powers [Figs.~\ref{fig8}(c) and (f)], the system exhibits increasingly complex dynamics with simultaneous coupling to multiple modes, resulting in broad energy redistribution across the spectrum.

Thus, nonlinearities can not only suppress BLZ oscillations, but also trigger new mode couplings. Moreover, BLZ oscillations exhibit remarkable robustness against weak nonlinear effects: the few-mode character of the dynamics is preserved, although the period and amplitude of oscillations can be noticeably altered. Notably, despite the broader range of instabilities typical in two-dimensional systems, no such instabilities are observed in our case.

\begin{figure}[t!]
\centering
\includegraphics[width=1.00\linewidth]{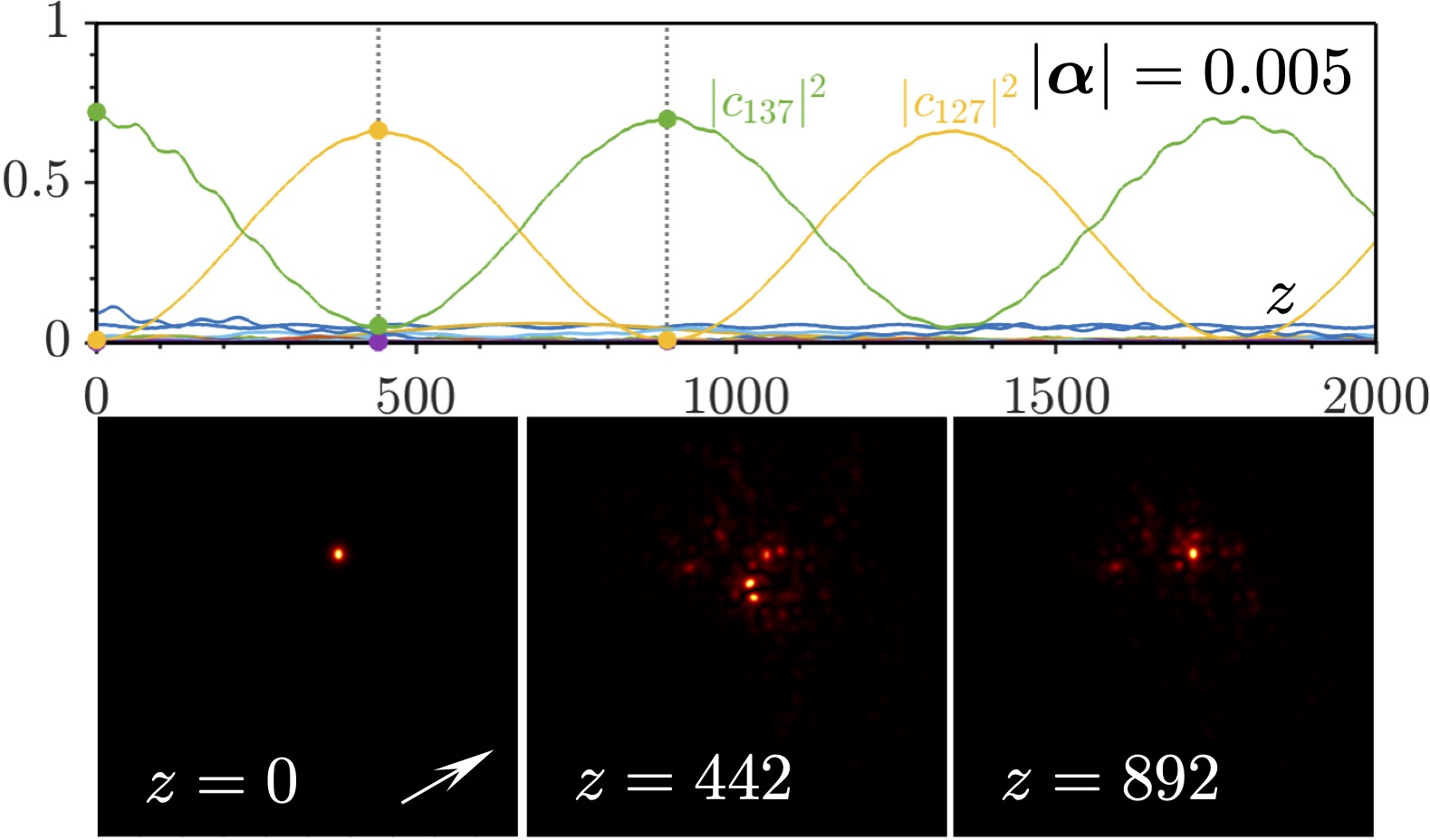}
\caption{{Evolution of weights $|c_n |^2$ of all linear modes of the system at $|\balpha|=0$ for the angle $\gamma=\pi/3$ and $|\balpha|=0.005$ (top) and field modulus distributions at different values of $z$ corresponding to the dots and vertical dashed lines in the top panels (bottom). Results are shown for linear propagation with a single-site excitation. The arrow indicates the direction of the gradient.}}
\label{fig9}
\end{figure}

\section{{Discussion}}

{
We now address several issues related to the experimental feasibility of observing two-dimensional BLZ oscillations. As mentioned in the Introduction, the system described by Eqs.~(\ref{main})-(\ref{pot_total}) can be realized in photorefractive crystals using method of optical induction \cite{Trompeter2006} that also allows creation of moir\'e potentials at will~\cite{Wang2020}. In such systems, transverse refractive index gradients can be realized by illumination of the sample with broad beam with linearly varying intensity. Similar configurations can also be created using highly developed technology of fs-laser inscription of waveguide arrays~\cite{Szameit2010} that was already successfully applied to creation of moire arrays \cite{Arkhipova2023}. In these systems, the transverse refractive index gradient can be emulated by bending of waveguides along the parabolic trajectory with distance $z$ \cite{Lenz1999, Dreisow2009}. 
}

{
In the last case, the dimensionless parameters used in the model are as follows. The transverse coordinates are defined by $\bm r=\tilde{\bm r}/r_0$, where $r_0=10$~$\mu\text{m}$ (hereafter tilde indicates the respective variables in the physical units) is the characteristic transverse scale and the dimensionless propagation distance $z=\tilde{z}/L_d$ is normalized to the diffraction length $L_d=kr_0^2\approx1.14$~$\text{mm}$, where $k=2\pi n/\lambda$ is the wavenumber at typical wavelength $\lambda=800$~$\text{nm}$, $n\approx1.45$ is the unperturbed refractive index of transparent dielectric material, where moir\'e lattice is inscribed (for example, fused silica). The dimensionless field amplitude is related to the real field amplitude $\mathcal{E}$ via $\Psi=\left(k^2r_0^2n_2/n\right)^{1/2}\mathcal{E}$, where the nonlinear refractive index (typical for fused silica) $n_2\approx2.7\times10^{-20}$~$\text{m}^2/\text{W}$. The dimensionless depth of sublattices forming moire lattice is given by $p_{1,2}=k^2r_0^2\delta n_{1,2}/n$, where $\delta n_{1,2}$ is the refractive index modulation depth in each sublattice. The representative values $p_{1,2}=4$ used here correspond to a refractive index modulation depth of $\delta n_{1,2}\approx4.5\times10^{-4}$. The dimensionless width of individual waveguides $w=0.5$ corresponds to $5$~$\mu\textrm{m}$, while the period $d=2.5$ of both sublattices corresponds $25$~$\mu\textrm{m}$. For experiments with low-power beams the samples with lengths up to $20-30$ cm can be fabricated, that correspond to dimensionless propagation distances of $z\approx180-270$.
}

{
It is relevant to notice that modern experimental techniques in optical waveguiding systems allow registration of not only output intensity distributions, but also of the intensity distributions at intermediate cross-sections inside the sample or even the entire light propagation dynamics inside the sample. Moreover, developed holographic techniques and spatial light modulators (SLMs) allow to create practically any desirable field distribution at the input face of the sample that can be focused into selected waveguides/location of the structure to excite a predetermined set of modes and to observe theoretically predicted dynamics.}

{Regarding the stability of BLZ oscillations, which is crucial for their experimental observation, deviations of the initial beam from the exact eigenmodes do not lead to instabilities, even in the weakly nonlinear regime, as demonstrated by the simulations above (obviously, no instabilities can arise in the linear case when modes do not interact with each other).
Concerning structural stability, i.e., deviations of the potential from the exact form used in the numerical simulations (illustrated in Sec.~\ref{subsec:perturbation}), no instability is observed due to the inherent nature of quasi-periodic systems and their discrete spectrum (for a finite specimen). Indeed, any small perturbation of  $V(\bm r)$ can only shift the discrete spectrum slightly and cause weak changes in the position and shape of the mode. However, such perturbations do not affect the selection rule itself. Thus, one may observe quantitative variations in the period and amplitude of the oscillations, but not their destruction.}

{{
Another important consideration is the preparation of the input signal, which depends on the experimental configuration. In the case of optical beams, the primary focus of this study, devices such as spatial light modulators enable the generation of nearly arbitrary input profiles. The programmability of a phase mask allows precise control over the spacing between excitation spots, their size, individual intensities, and relative phases. Although various methods exist to create specific input conditions, one of the simplest approaches is a single-site excitation. To assess the robustness and experimental feasibility of BLZ oscillations, we consider the parameters $\gamma = \pi/3$, $|\alpha| = 0.005$, and $\theta = \pi/3$, with a single-site excitation at the waveguide corresponding to the maximum intensity of the mode $n = 137$. This mode served as the initial condition for the dynamics depicted in Fig.~\ref{fig5}(b). The results of the single-site excitation are presented in Fig.~\ref{fig9}. Although the oscillation amplitudes of the mode weights $|c_n|^2$ are slightly reduced due to the radiation, both the oscillation period and the modal composition remain almost unchanged relative to the case of mode-driven excitation. These results demonstrate that BLZ oscillations can be reliably excited in an optical system using a single-beam input.
}}

\section{Conclusion} 

To conclude, we have developed a theory of two-dimensional BLZ oscillations for localized modes in aperiodic lattices, illustrating the results with a light beam propagating in the incommensurate moiré lattice with a transverse refractive index gradient. {Since modes with propagation constants (energies) above (below) the mobility edge are localized in the coordinate space but possess different propagation constants (energies), the linear dynamics is governed by the simultaneous spatial tunneling and Landau-Zener tunneling in the energy space.} This dynamics is highly sensitive to the initial conditions and may involve several resonantly coupled modes determined by the selection rule. {In the 2D setting, in contrast to their 1D counterpart, the selection rule depends on the angle of the gradient, which can result in resonantly coupled modes being misaligned with the direction of the linear gradient.} {The dependence on the gradient arises from the spatial localization of the modes, which further explains why quasi-resonances at avoided crossings are more relevant for the dynamics than the level crossings that are typical in 2D settings.} Furthermore, BLZ oscillations may not appear at small gradients or become suppressed with the increase of the gradient, if no modes satisfying the selection rule are found. Weak nonlinearity, {whether attractive or repulsive,} preserves the few-mode character of the dynamics, although it can significantly modify the period and amplitude of the oscillations.  

The reported theory allows direct generalization, including different types of aperiodic potentials, like quasicrystals, or described by two-dimensional almost-periodic functions, {all of which have been successfully realized in laboratory settings}. 

Although our analysis focuses on optical systems, the underlying theory is directly applicable to other experimentally available settings such as BECs~{\cite{Meng2023}, where optical lattices in general are highly tunable~\cite{Jo2012,Taie2015}},  or acoustic waves~{~\cite{Han2024}} in systems governed by moir\'e, {or more generally, quasi-periodic} potentials.



\medskip

\section*{Acknowledgements}

SKI has received funding from the European Union through the Program Fondo Social Europeo Plus 2021-2027(FSE+) of the Valencian Community (Generalitat Valenciana CIAPOS/2023/329). YVK was supported by the Russian Science Foundation (grant 24-12-00167) and partially by the research project FFUU-2024-0003 of the Institute of Spectroscopy of the Russian Academy of Sciences. VVK was supported by the Fundação para a Ciência e Tecnologia under the projects 2023.13176.PEX (DOI \\https://doi.org/10.54499/2023.13176.PEX) and by national funds, under the Unit CFTC - Centro de Física Teórica e Computacional, reference UID/00618/2023, financing period 2025-2029.

\section*{Data availability statement}
All data that support the findings of this study are included within the article or available upon reasonable request from the authors.

\section*{Conflict of Interest}
The authors declare no conflict of interest

\medskip


\begin{thebibliography}{99}
 	
 	\bibitem{Bloch1929} F. Bloch, 
 	{\it Z. Phys.} {\bf 1929}, {\it 52}, 555.
   	
 	
 	\bibitem{Mendez1988}  E.~E.~Mendez,  F.~Agull\'o-Rueda,   J.~M.~Hong,
 	{\it Phys. Rev. Lett.} {\bf 1988}, {\it 60}, 2426.
 	
 	\bibitem{FeLeSh} J.~Feldmann, K.~Leo, J.~Shah, D.~A.~B.~Miller, J.~E.~Cunningham,  T.~Meier, G.~von~Plessen, A.~Schulze, P.~Thomas, S.~Schmitt-Rink,
 	{\it Phys. Rev. B} {\bf 1992}, {\it 46}, 7252(R).
 	
 	\bibitem{Leo1992} K.~Leo, P.~H.~Bolivar,  F.~Br\"uggemann, R.~Schwedler, K.~K\"ohler,
 	{\it Solid State Commun.} {\bf 1992}, {\it 84}, 943.
 	
 	\bibitem{Waschke1993}
    C.~Waschke,   H.~G.~Roskos, R.~Schwedler, K.~Leo, H.~Kurz, K.~K\"ohler, 
 	{\it Phys. Rev. Lett.} {\bf 1993}, {\it 70}, 3319.
 	
 	
 	
 	
 	\bibitem{Dahan1996}   M.~B.~Dahan, E.~Peik, J.~Reichel, Y.~Castin, C.~Salomon, 
 	{\it Phys. Rev. Lett.} {\bf 1996}, {\it 76}, 4508.

{
\bibitem{Wilkinson1996} S. R. Wilkinson, C. F. Bharucha, K. W. Madison, Qian Niu, and M. G. Raizen,
{\it Phys. Rev. Lett.} {\bf 1996} {\it 76}, 4512. 
}

    
 	\bibitem{AndKas}  B.~P.~Anderson, M.~A.~Kasevich 
 	{\it Science} {\bf 1998}, {\it 282}, 1686.
 	
 	\bibitem{Morsch2001} O.~Morsch,   J.~H.~M\"uller, M.~Cristiani, D.~Ciampini, 
 	E.~Arimondo, 
 	{\it Phys. Rev. Lett.} {\bf 2001} {\it 87}, 140402.
 	
 	\bibitem{Cristiani2002} M.~Cristiani, O.~Morsch, J.~H.~M\"uller, D.~Ciampini, E.~Arimondo,
 	{\it Phys. Rev. A} {\bf 2002}, {\it 65}, 063612.
 	
 	\bibitem{Ferrari2006} G.~Ferrari, N.~Poli, F.~Sorrentino,  G.~M.~Tino,
 	{\it Phys. Rev. Lett.} {\bf 2006}, {\it 97}, 060402.
 	 
 	\bibitem{Gustavsson2008} M.~Gustavsson, E.~Haller,  M.~J.~Mark,  J.~G.~Danzl, G.~Rojas-Kopeinig,  H-C.~N\"agerl, 
 	{\it Phys. Rev. Lett.} {\bf 2008}, {\it 100}, 080404.
 	 	
 	\bibitem{Kling2010} S.~Kling, T.~Salger, C.~Grossert,  M.~Weitz,
 	{\it Phys. Rev. Lett.} {\bf 2010}, {\it 105}, 215301.
 	
 	\bibitem{Geiger2018}  Z.~A.~Geiger,
 	K.~M.~Fujiwara, K.~Singh, R.~Senaratne, S.~V.~Rajagopal, M.~Lipatov, T.~Shimasaki, R.~Driben, V.~V.~Konotop, T.~Meier, D.~M.~Weld,
 	{\it Phys. Rev. Lett.} {\bf 2018}, {\it 120}, 213201.
 	
   	
 	\bibitem{Peschel1998}  U.~Peschel, T.~Pertsch, F.~Lederer, 
 	{\it Opt. Lett.} {\bf 1998}, {\it 23}, 1701.
 	
 	\bibitem{Pertsch1999} T.~Pertsch, P.~Dannberg, W.~Elflein, A.~Br\"auer, F.~Lederer, 
 	Phys. Rev. Lett. {\bf 1999}, {\it 83}, 4752.
 	
 	\bibitem{Morandotti1999}  R.~Morandotti, U.~Peschel, J.~S.~Aitchison, H.~S.~Eisenberg, Y.~Silberberg, 
 	{\it Phys. Rev. Lett.} {\bf 1999}, {\it 83}, 4756.
 
 	\bibitem{Sapienza2003} R.~Sapienza, P.~Costantino, D.~Wiersma, M.~Ghulinyan, C.~J.~Oton, L.~Pavesi,
 	{\it Phys. Rev. Lett.} {\bf 2003}, {\it 91}, 263902.
 
 	\bibitem{Dreisow2009} F.~Dreisow, A.~Szameit, M.~Heinrich, T.~Pertsch, S.~Nolte, A.~T\"unnermann, S.~Longhi,
 	{\it Phys. Rev. Lett.} {\bf 2009}, {\it 102}, 076802.
  
   	\bibitem{Yuan2016} L.~Yuan, S.~Fan, 
 	{\it Optica} {\bf 2016}, {\it 3}, 1014.

 
 		
 		
 	\bibitem{Reeves2014}  J.~B.~Reeves, B.~Gadway, T.~Bergeman, I.~Danshita, D.~Schneble, 
 	{\it New J. Phys.} {\bf 2014}, {\it 16}, 065011.

 	
 	\bibitem{Luschen2018}  H.~P.~L\"uschen, S.~Scherg, T.~Kohlert,  M.~Schreiber, P.~Bordia, X.~Li, S.~Das~Sarma,  I.~Bloch,
 	{\it Phys. Rev. Lett.} {\bf 2018}, {\it 120}, 160404.
 	  	 
 	
 	\bibitem{Kraus2012} Y E.~Kraus, Y.~Lahini, Z.~Ringel, M.~Verbin, O.~Zilberberg,
 	 	{\it Phys. Rev. Lett.} {\bf 2012}, {\it 109}, 106402.
 	 
  
 	 \bibitem{Yang2024} 
 	 K.~Yang, Q.~Fu, H.~Prates, C.~Huang, P.~Wang, Y.~V.~Kartashov, V.~V.~Konotop, F.~Ye, 
 	 \textit{Proc. Natl Acad. Sci. USA} {\bf 2024}, {\it 121} e2411793121.
 		
 
 \bibitem{Aubri1980} S.~Aubry, G.~Andr\'e,
 \textit{Ann. Isr. Phys. Soc.} {\bf 1980}, {\it 3}, 133.
 
 
  
 \bibitem{Moura2005} F.~A.~B.~F.~de~Moura, M.~L.~Lyra, F.~Dom\'inguez-Adame, V.~A.~Malyshev,  
 {\it Phys. Rev. B} {\bf 2005}, {\it 71}, 104303.
 
 \bibitem{Wang2014} G.~Wang, 
  {\it J. Opt.} {\bf 2014}, {\it 16}, 015502.

 \bibitem{Walter2010} S.~Walter, D.~Schneble, A.~C.~Durst, 
   {\it Phys. Rev. A} {\bf 2010}, {\it 81}, 033623.

\bibitem{Prates2024}   H.~C.~Prates, V.~V.~Konotop, 
{\it Phys. Rev. Res.} {\bf 2024}, {\it 6}, L022011.
   
   \bibitem{Simon1982} B.~Simon,
   {\it Adv. App. Math.} {\bf 1982}, {\it 3}, 463.
  
 
  
  \bibitem{Sarnak1982} P.~Sarnak, 
  {\it Comm. Math. Phys.} {\bf 1982}, {\it 84}, 377.
   
 \bibitem{Kohmoto83} M.~Kohmoto,
 {\it Phys. Rev. Lett.} {\bf 1983}, {\it 51}, 1198.
 
 \bibitem{Surace1990} S.~Surace,~Jr.,
 {\it Trans. Am. Math. Soc.} {\bf 1990}, {\it 320}, 321.
 
 \bibitem{Diener01} R.~B.~Diener, G.~A.~Georgakis, J.~Zhong, M.~Raizen, Q.~Niu, 
 {\it Phys. Rev. A} {\bf 2001}, {\it 64}, 033416.
  
 \bibitem{Modugno2009}  M.~Modugno, 
 {\it New J. Phys.} {\bf 2009}, {\it 11}, 033023.
 
 \bibitem{Biddle2002} J.~Biddle, B.~Wang, D.~J.~Priour, S.~Das~Sarma, 
 {\it Phys. Rev. A} {\bf 2009}, {\it 80}, 021603(R).


 
 
 
 	\bibitem{Zener1932} C.~Zener,
 {\it Proc. R. Soc. Lond. Ser. A.} {\bf 1932}, {\it 137}, 696.
 
 \bibitem{LandauTun1932} L.~D.~Landau, 
 {\it Phys. Z. Sowjet.} {\bf 1932}, {\it 1}, 88. 


{\bibitem{Sias2007} C. Sias, A. Zenesini, H. Lignier, S. Wimberger, D. Ciampini, O. Morsch, and E. Arimondo,
{\it Phys. Rev. Lett.} {\bf 2007} {\it 98}, 120403. 
\bibitem{Wilkinson1997} S.~R.~Wilkinson, C.~F.~Bharucha,
M.~C.~Fischer, K.~W.~Madison, P.~R.~Morrow, Q.~Niu, B.~Sundaram, 
M.~G.~Raizen {\it Nature}, {\bf 1997} {\it 387}, 575.
%
  \bibitem{Zenesini2008} A.~Zenesini, C.~Sias, H.~Lignier, 
Y.~Singh, D.~Ciampini, O.~Morsch, R.~Mannella, E.~Arimondo, A.~Tomadin, S.~Wimberger,
{\it New J. Phys.} {\bf 2008} {\it 10}, 053038
%
\bibitem{Zenesini2009} 
A.~Zenesini, H.~Lignier, G.~Tayebirad, J.~Radogostowicz, D.~Ciampini, R.~Mannella, S.~Wimberger, O.~Morsch, E.~Arimondo,
Phys. Rev. Lett. {\bf 2009}, {\it 103}, 090403.
%
\bibitem{Tayebirad2010}
G.~Tayebirad, A.~Zenesini, D.~Ciampini, R.~Mannella, O.~Morsch, E.~Arimondo, N.~L\"orch, S.~Wimberger,
{\it Phys. Rev. A} {\bf 2010} {\it 82}, 013633.
}
 



 
 
 \bibitem{Gluck2001} M.~Gl\"uck, F.~Keck,  A.~R.~Kolovsky, H.~J.~Korsch,
 {\it Phys. Rev. Lett.}  {\bf 2001}, {\it 86}, 3116.
 
 \bibitem{Kolovsky2003}   A.~R.~Kolovsky,   H.~J.~Korsch, 
 {\it Phys. Rev. A} {\bf 2003}, {\it 67}, 063601.

\bibitem{Witthaut2004}  D.~Witthaut, F.~Keck,  H.~J.~Korsch,  S.~Mossmann,
 {\it New J. Phys.} {\bf 2004}, {\it 6},  41.

\bibitem{Mossmann2005} S.~Mossmann, A.~Schulze, D.~Witthaut,   H.~J.~Korsch,
 {\it J. Phys. A: Math. Gen.} {\bf 2005}, {\it 38}, 3381.
 
 \bibitem{Khomeriki2010}  R.~Khomeriki,
 {\it Phys. Rev. A}  {\bf 2010}, {\it 82}, 033816.

{
\bibitem{Kolovsky2011} A.~R.~Kolovsky, G. Mantica,  Phys. Rev. E 83, 041123 (2011).
{\it Phys. Rev. E} {\bf 2011}, 83, 041123. 
\bibitem{Kolovsky2012_1} A.~R.~Kolovsky, E.~N.~Bulgakov, {\it Front. Phys.} {\bf 2012} {\it 7}, 3. 
\bibitem{Kolovsky2012_2} A.~R.~Kolovsky, 
{\it Phys. Rev. E} {\bf 2012}, {\it 86}, 041146 

\bibitem{Kolovsky2013} A.~R.~Kolovsky, E.~N.~Bulgakov,
{\it Phys. Rev. A} {\bf 2013}, {\it 87}, 033602.
\bibitem{Maksimov2015} 
D.~N.~Maksimov, E.~N.~Bulgakov, A.~R.~Kolovsky
{\it Phys. Rev. A}, {\bf 2015} 91, 053632.
\bibitem{Kolovsky2016} A.~R.~Kolovsky,
{\it Phys. Rev. A}, {\bf 2016}, {\it 93}, 033626.
\bibitem{Driben2017} R.~Driben, V.~V.~Konotop, T.~Meier, and A.~V.~Yulin,
{\it Sci. Rep.} {\bf 2017}, {\it 7}, 3194.
} 

  
 \bibitem{Ye2023} L.-L.~Ye, Y.-C.~Lai,
 {\it Phys. Rev. B} {\bf 2023}, {\it 107}, 165422.
 
 
 \bibitem{Trompeter2006} H.~Trompeter, W.~Krolikowski, D.~N.~Neshev,   A.~S.~Desyatnikov,   A.~A.~Sukhorukov,
 Y.~S.~Kivshar, T.~Pertsch, U.~Peschel, F.~Lederer, 
 {\it Phys. Rev. Lett.} {\bf 2006}, {\it 96}, 053903.

  
 \bibitem{He2007} Z.~He, S.~Peng, F.~Cai, M.~Ke, Z.~Liu,
 {\it Phys. Rev. E} {\bf 2007}, {\it 76}, 056605.
  
 
 \bibitem{Moura2008} F.~A.~B.~F.~de~Moura, L.~P.~Viana, M.~L.~Lyra,  V.~A.~Malyshev, F.~Dom\'inguez-Adame, 
 {\it Phys. Lett. A} {\bf 2008}, {\it 372}, 6694.
 
 \bibitem{Yar2023} A.~Yar, B.~Sarwar, S.~B.~A.~Shah, K.~Sabeeh,
 {\it Phys. Lett. A} {\bf 2023}, {\it 478}, 128899.
 
 
 
 \bibitem{Wang2020} P.~Wang, Y.~Zheng, X.~Chen, C.~Huang, Y.~V.~Kartashov, L.~Torner, V.~V.~Konotop,  F.~Ye,
 {\it Nature} {\bf 2020}, {\it 577}, 42.
 
 \bibitem{Fu2020} Q.~Fu,  P.~Wang, C.~Huang,  Y.~V.~Kartashov, L.~Torner, V.~V.~Konotop,  F.~Ye, 
 {\it Nat. Phot.} {\bf 2020}, {\it 14}, 663.
   
 \bibitem{Wang2022} P.~Wang, Q.~Fu, R.~Peng, Y.~V.~Kartashov, L.~Torner, V.~V.~Konotop, F.~Ye,
{\it Nat. Commun.} {\bf 2022}, {\it 13}, 6738.
 
 \bibitem{Arkhipova2023} A.~A.~Arkhipova, Y.~V.~Kartashov, S.~K.~Ivanov, S.~A.~Zhuravitskii, N.~N.~Skryabin, I.~V.~Dyakonov, A.~A.~Kalinkin, S.~P.~Kulik, V.~O.~Kompanets, S.~V.~Chekalin, F.~Ye, V.~V.~Konotop, L.~Torner, V.~N.~Zadkov,
{\it Phys. Rev. Lett.} {\bf 2023}, {\it 130}, 083801.
 
 
 \bibitem{Hurley2023} N.~Hurley,  S.~Kamau, J.~Cui, Y.~Lin,
 {\it Micromachines} {\bf 2023}, {\it 14}, 1217.
  
 
\bibitem{Shang2021} C.~Shang, C.~Lu, S.~Tang, Y.~Gao, Z.~Wen,
 {\it Opt. Expr.} {\bf 2021}, {\it 29}, 29116.

 \bibitem{Toyoda1983}
 T.~Toyoda, M.~Yabe,
 {\it J. Phys. D} {\bf 1983}, {\it 16}, L97.

\bibitem{Plotnik2011}
Y.~Plotnik, O.~Peleg, F.~Dreisow, M.~Heinrich, S.~Nolte, A.~Szameit, M.~Segev,
 {\it Phys. Rev. Lett.} {\bf 2011}, {\it 107}, 183901.
 

\bibitem{Avron1977} J.~E.~Avron, J.~Zak, A.~Grossmann, L.~Gunther,
{\it J. Math. Phys.} {\bf 1977}, {\it 18}, 918.

\bibitem{Bentosela1983} F.~Bentosela, R.~Carmona, P.~Duclos, B.~Simon, B.~Souillard, R.~Weder,
{\it Commun. Math. Phys.} {\bf 1983}, {\it 88}, 387.


\bibitem{Nenciu1991} G.~Nenciu,
{\it Rev. Mod. Phys.} {\bf 1991}, {\it 63}, 91.

\bibitem{Huang2016}
C.~Huang, F.~Ye, X.~Chen, Y.~V.~Kartashov, V.~V.~Konotop, L.~Torner, 
{\it Sci. Rep.} {\bf 2016}, {\it 6}, 32546.

\bibitem{Prates2022} H.~C.~Prates, D.~A.~Zezyulin, V.~V.~Konotop, 
{\it Phys. Rev. Res} {\bf 2022}, {\it 4}, 033219.

 

\bibitem{Konotop2024} V.~V.~Konotop,
{\it Phys. Rev. Res.} {\bf 2024}, {\it 6}, 033113.
 
 
 
 
 
 
 
 	

 		 
 		 
 
 	 	
 	
  	
\bibitem{Gershgorin} S.~Gershgorin,
 	{\it Izv. Akad. Nauk. USSR Otd. Fiz.-Mat. Nauk} {\bf 1931}, {\it 6}, 749.

 

{
\bibitem{Szameit2010}
A. Szameit, S. Nolte,
\textit{J. Phys. B: At. Mol. Opt. Phys.} \textbf{2010}, \textit{43}, 163001.
\bibitem{Lenz1999}
G. Lenz, I. Talanina, C. Martijn de Sterke,
\textit{Phys. Rev. Lett.} \textbf{1999}, \textit{83}, 963.
\bibitem{Meng2023} Z.~Meng, L.~Wang, W.~Han, F.~Liu, K.~Wen, C.~Gao, P.~Wang, C.~Chin, J.~Zhang, 
{\it Nature} {\bf 2023}, {\it 615} 231.
\bibitem{Jo2012} G.-B.~Jo, J.~Guzman, C.~K.~Thomas, P.~Hosur, A.~Vishwanath, D. M. Stamper-Kurn, 
{\it Phys. Rev. Lett.} {\bf 2012}, {\it 108}, 045305.
\bibitem{Taie2015} S.~Taie, H.~Ozawa, T.~Ichinose, T.~Nishio, S.~Nakajima, Y. Takahashi, 
{\it Sci. Adv.} {\bf 2015}, {\it 1}, e1500854.
\bibitem{Han2024} C.~Han, L.~Q.~~Chen, T.~Yang,  G.~Xu, J.~Li, C.~Li, H.~Fan, A.~Al\`{u}, C.-W. Qiu,
{\it Nat. Commun.} {\bf 2025}, {\it 16}, 1988. 
} 	

 	
 	
  
   
 	
 	
 \end{thebibliography}
\end{document}